\documentclass[10pt, final, journal]{IEEEtran}

\usepackage{authblk}

\usepackage{cite}

\usepackage{booktabs}
\usepackage{color}

\usepackage{threeparttable}
\usepackage{multirow}
\usepackage{tabularx}
\usepackage{threeparttable}
\usepackage{makecell}
\ifCLASSINFOpdf
  \usepackage[pdftex]{graphicx}
  \usepackage{subfigure}
  \graphicspath{{../pdf/}{../jpeg/}}
\else
  
\fi

\usepackage{amsmath}
\usepackage{amsfonts}

\usepackage{algorithmic}

\usepackage{array}

\begin{document}

\title{Temporal-Framing Adaptive Network \\ for Heart Sound Segmentation without \\ Prior Knowledge of State Duration}

\author{
    \IEEEauthorblockN{Xingyao Wang,
                      Chengyu Liu\IEEEauthorrefmark{1}, ~\IEEEmembership{Senior~Member, ~IEEE},
                      Yuwen Li,
                      Xianghong Cheng\IEEEauthorrefmark{1},
                      Jianqing Li
                      and Gari D. Clifford, ~\IEEEmembership{Senior~Member, ~IEEE}%
                      \thanks{X. Y. Wang, C. Y. Liu, Y. W. Li, X. H. Cheng and J. Q. Li are with the School of Instrument Science and Engineering, Southeast University, Nanjing, 210096, China. X. Y. Wang and C. Y. Liu are also with the State Key Laboratory of Bioelectronics, Southeast University, Nanjing, 210096, China (e-mails: chengyu@seu.edu.cn and xhcheng@seu.edu.cn).}%
                      \thanks{G. D. Clifford is with Department of Biomedical Informatics, Emory University School of Medicine, Atlanta, GA 30322, USA, as well as with Department of Biomedical Engineering, Georgia Institute of Technology, Atlanta, GA 30332, USA.}
                      }
}
\maketitle
\footnote{Copyright (c) 2020 IEEE. Personal use of this material is permitted. However, permission to use this material for any other purposes must be obtained from the IEEE by sending an email to pubs-permissions@ieee.org. This paper is an accepted version for publication in IEEE Transactions on Biomedical Engineering.}
\begin{abstract}
\emph{Objective:} This paper presents a novel heart sound segmentation algorithm based on Temporal-Framing Adaptive Network (TFAN), including state transition loss and dynamic inference.
\emph{Methods:} In contrast to previous state-of-the-art approaches, TFAN does not require any prior knowledge of the state duration of heart sounds and is therefore likely to generalize to non sinus rhythm. 
TFAN was trained on 50 recordings randomly chosen from Training set A of the 2016 PhysioNet/Computer in Cardiology Challenge and tested on the other 12 independent databases (2,099 recordings and 52,180 beats).
And further testing of performance was conducted on databases with three levels of increasing difficulty (LEVEL-I, -II and -III).
\emph{Results:} TFAN achieved a superior \(F_1\) score for all 12 databases except for `Test-B', with an average of 96.72\%, compared to 94.56\% for logistic regression hidden semi-Markov model (LR-HSMM) and 94.18\% for bidirectional gated recurrent neural network (BiGRNN).
Moreover, TFAN achieved an overall \(F_1\) score of 99.21\%, 94.17\%, 91.31\% on LEVEL-I, -II and -III databases respectively, compared to 98.37\%, 87.56\%, 78.46\% for LR-HSMM and 99.01\%, 92.63\%, 88.45\% for BiGRNN. 
\emph{Conclusion:} TFAN therefore provides a substantial improvement on heart sound segmentation while using less parameters compared to BiGRNN.
\emph{Significance:} The proposed method is highly flexible and likely to apply to other non-stationary time series. Further work is required to understand to what extent this approach will provide improved diagnostic performance, although it is logical to assume superior segmentation will lead to improved diagnostics. 
\end{abstract}

\begin{IEEEkeywords}
Heart sound segmentation, phonocardiogram, deep neural networks, hidden semi-Markov models.
\end{IEEEkeywords}

\section{Introduction}

\IEEEPARstart{C}{ardiac} auscultation, for identifying heart sounds, is commonly the first step and the most cost-effective measure for screening the various heart dysfunction, even though the final diagnosis is based on the combined analysis from a series of electrophysiologic study or ultrasound recordings. Heart sounds can reflect the hemodynamic processes of the heart and identify some representative symptoms of different diseases, including arrhythmia, valve disease, pulmonary hypertension, heart failure, among other issues \cite{tavel1996cardiac}. However, only about \(20\%\) of medical interns can effectively detect heart conditions using auscultation\cite{mangione1993teaching}, and extensive training is necessary for human expert evaluation. Automatic and accurate analysis of the recording of heart sounds (the phonocardiogram, or PCG) can be useful for auxiliary diagnosis in clinical applications, and it can potentially assist interns with less developed skills.

The segmentation of heart sounds is a critical step in the automatic analysis of PCG. Accurate localization of fundamental components in PCG is a pre-condition of mining more specific pathological information, including the preliminary diagnosis of specific pathogenic sites and severity levels of these heart diseases \cite{liu2016open}. Although unsupervised approaches can facilitate classification or prediction, the lack of interpretability is likely to be a significant barrier to clinical adoption. 

\begin{figure}[!t]
  \centering
    \includegraphics[width=3.83in]{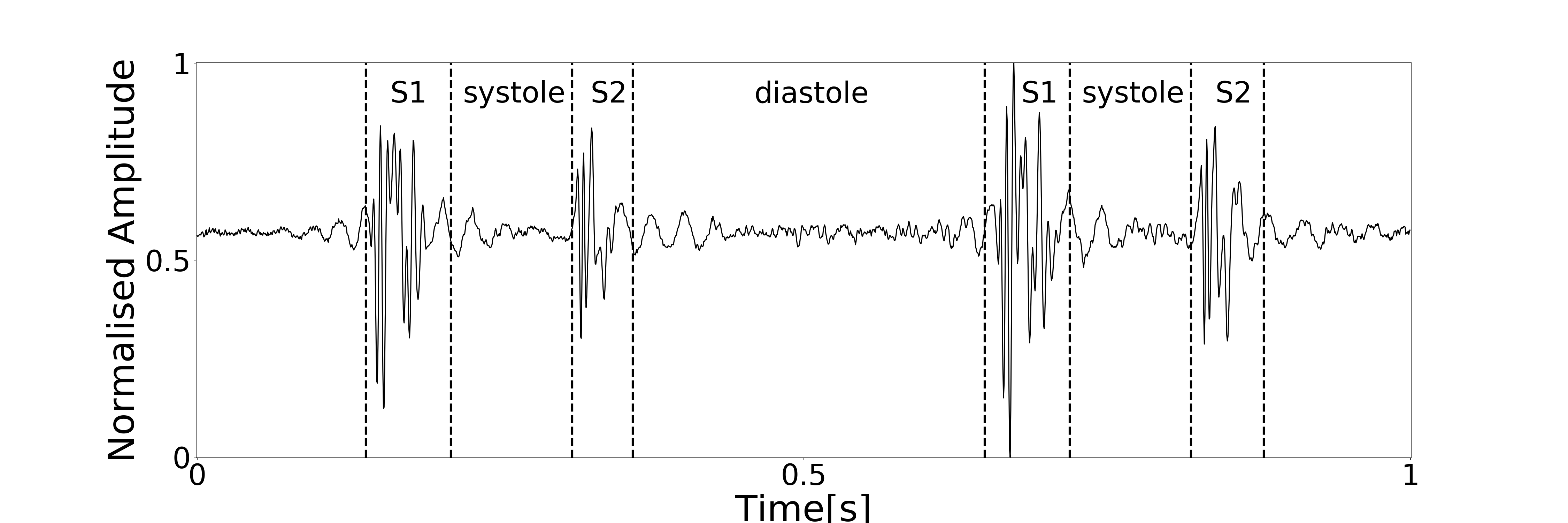} \\
  \caption{Example of a recorded PCG signal and four states of the heart cycle (S1, systole, S2, diastole).}
  \label{fig1}
\end{figure}

Each heart cycle usually  consists of a sequence of temporally-constrained states; the first heart sound (S1), the systolic period, the second heart sound (S2) and then the  diastolic period (Fig.\ref{fig1}). Segmentation of the PCG into these states facilitates further (pathological) feature extraction within different periods of each heart cycle, e.g., the audible third and fourth heart sound (S3 and S4), murmurs, ejection clicks, pericardial ``knock'', \emph{etc}. In addition, segmentation into these states allows for the detection of abnormalities in the timing of different sounds. For example, a mid or late systolic click is most likely a diagnostic indicator of mitral (or tricuspid) valve prolapse, even though echocardiograms may fail to confirm this finding \cite{tavel1996cardiac}.

In earlier works, segmentation of heart sounds have leveraged features from both the time and frequency domain \cite{carvalho2005low, vepa2008segmentation, pedrosa2014automatic, ari2008robust, yan2010moment}, including: Shannon energy \cite{liang1997heart}, wavelet envelope \cite{huiying1997heart}, Hilbert transform \cite{sun2014automatic}, time-frequency transform \cite{moukadem2013robust, castro2013heart, naseri2013detection}, and cepstral coefficients \cite{vepa2009classification, kumar2006new, chen2016s1}, \emph{etc}. These features have been used both directly (on a sliding window) or to generate observable sequences from heart sounds for probabilistic sequence models, like hidden Markov models (HMMs) and their variations \cite{springer2015logistic, oliveira2018adaptive, noman2019markov}. Although HMM-based methods have the advantage of modeling the sequential and periodic nature of heart sounds, this can result in false positives when noise or artifacts occur with similar features to the heart sounds. 
To mitigate this problem, Gill \emph{et al.} \cite{gill2005detection} proposed the incorporation of timing durations within HMM for heart sound segmentation. Schmidt \emph{et al.} \cite{schmidt2010segmentation} were the first to explicitly model the expected duration of heart sounds within the HMM framework using a hidden semi-Markov model (HSMM). Springer \emph{et al.} \cite{springer2015logistic} extended this work by modifying the Viterbi algorithm to include the duration densities and adding a logistic regression emission layer. 
This logistic regression-HSMM-based (LR-HSMM) method was evaluated on 10,172 seconds of heart sounds collected from 112 (healthy and pathological) subjects (with simultaneous electrocardiogram (ECG) as a gold standard) and demonstrated an average \(F_1\) score of \(98.5\%\) for segmenting S1 and \(97.2\%\) for segmenting S2 \cite{liu2017performance}. This method was adopted as the reference segmentation method in the 2016 PhysioNet/Computet in Cardiology (CinC) Challenge for the classification of normal and abnormal heart sounds \cite{clifford2017recent}.

Nevertheless, the dependence of prior knowledge of state duration makes the method prone to false negatives during arrhythmia, particularly for tachycardia and bradycardia. 
As reported by the authors in the assessment of the 2016 PhysioNet/CinC Challenge \cite{clifford2017recent}, Not all researchers adopted Springer's segmentation method as the first step in the required classification task \cite{potes2016ensemble, kay2017dropconnected, zabihi2016heart, yang2016classification}. 
Whereas, the accuracy of anomaly recognition was not distinguished by this.
It is indicated that segmentation does not necessarily result in an obvious improvement in classification.
And further development of the segmentation algorithm may result in superior performance.

Recently, various approaches for reducing the explicit restrictions on state duration in heart sound segmentation have been proposed. Messner \emph{et al.} \cite{messner2018heart} suggested an event detection approach using bidirectional gated recurrent neural network (BiGRNN) and achieved a similar performance compared to Springer's method. 
Renna \emph{et al.} \cite{renna2019deep} utilized 1D U-Net \cite{ronneberger2015u} as transformation for HMMs and HSMMs. Meanwhile, with the progress in convolution neural network (CNN) for temporal data \cite{rethage2018wavenet, peddinti2017low}, it seems obvious to apply these techniques in this context. However, such deep-learning-based (DL-based) approaches are known to overfit on the differences in noise levels between databases, due to recording conditions and device heterogeneity. 

In this work, we propose an algorithm that combines both automated feature learning and sequential modeling. In order to eliminate the instability of the segmentation on pathological and noisy PCG signals, the proposed method disuses prior knowledge of heart sound state duration. The main contributions of this paper are:

1) Designing an adaptive Wiener filter to reduce the variabilities of the characteristics from different stethoscopes on heart sounds.

2) Developing an adaptive learning method to detect the four states of heart sounds, including building a temporal-framing adaptive network (TFAN) for the frame-level recognition, and designing state transition loss and dynamic inference.

3) Testing and comparing the proposed method with two state-of-the-art methods, the LR-HSMM method \cite{springer2015logistic} and the BiGRNN-based method \cite{messner2018heart}, over the whole database from PhysioNet/CinC Challenge 2016 and data sets with different segmentation difficulties. 

\begin{figure*}[!ht]
  \centering
  \begin{minipage}[b]{6.8in}
    \centering
    \includegraphics[width=6.8in]{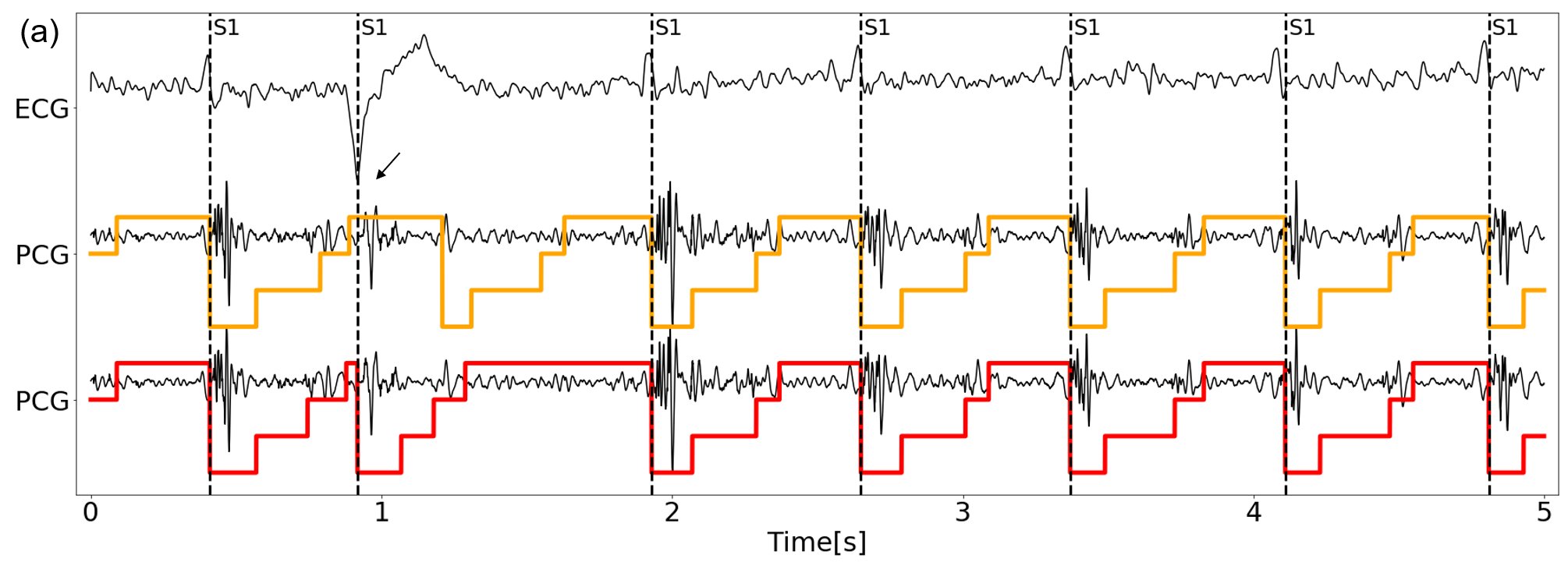} \\
    \includegraphics[width=6.8in]{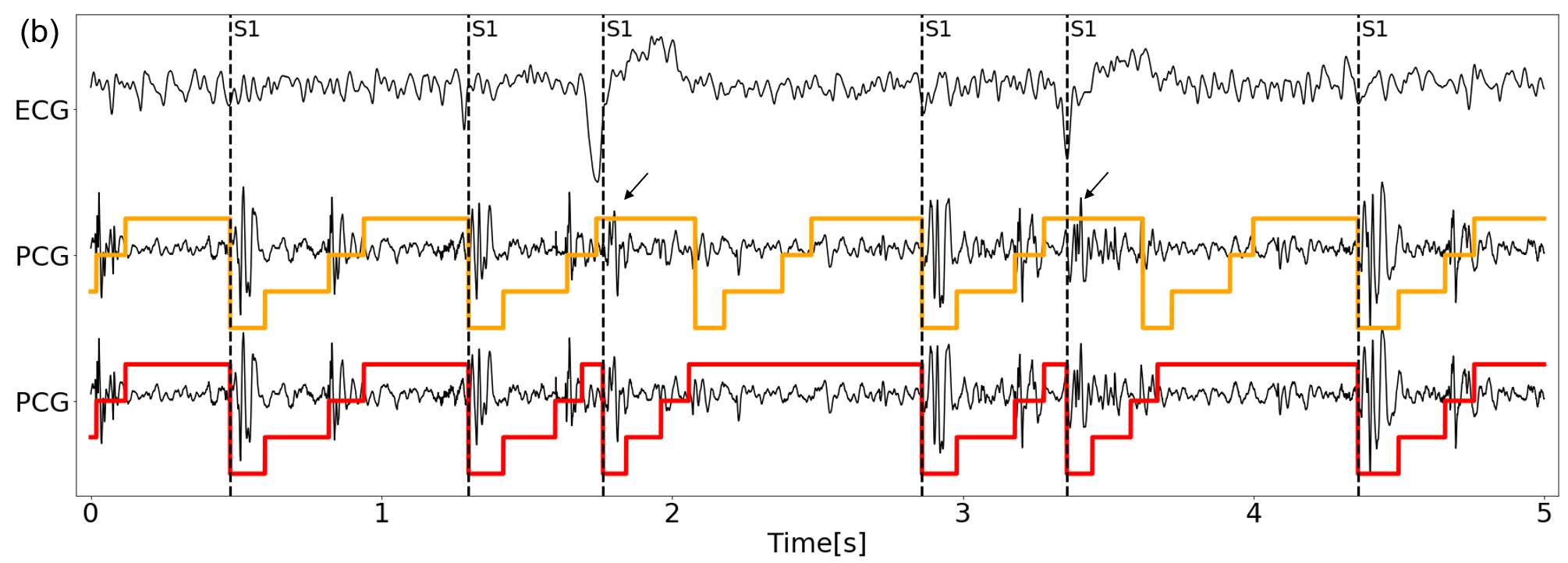} \\
    \includegraphics[width=6.8in]{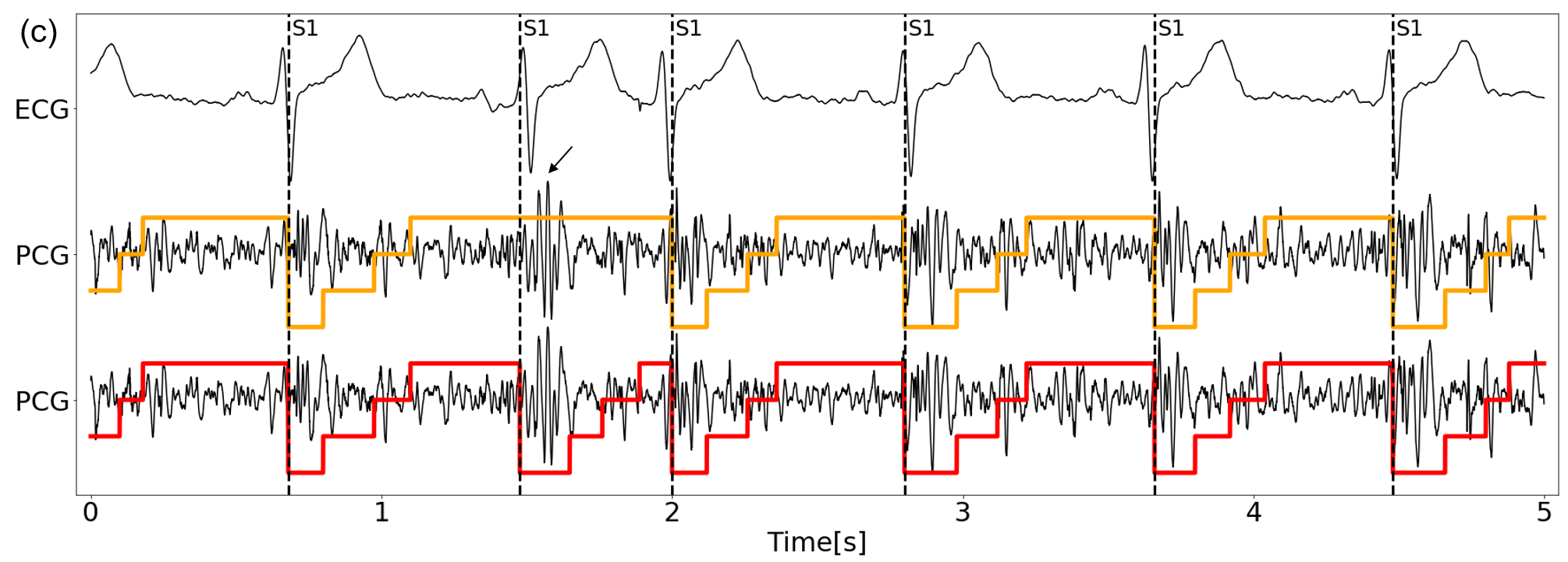}
  \end{minipage}
  \caption{Illustrations of the ECG and simultaneous PCG with automatically derived states in Training-A. Sub-figures (a) and (b) are from patients with premature ventricular contractions (PVCs), and sub-figure (c) is from a patient with premature atrial contractions (PACs). Arrhythmia-like PACs or PVCs always induce the false annotations in PCG signals (middle waveform in yellow) and the arrows point out the mistakes in the original annotation. The corrected annotations are shown in each sub-figure as a repeated waveform and a red staircase plot overlaid. Each level in the staircase plot represents S1, systole, S2 and diastole (in ascending value).}
  \label{fig2}
\end{figure*}

\section{DATABASE}
\subsection{General Introduction}
The 2016 PhysioNet/CinC Challenge \cite{clifford2016classification} contains 12 independent data sets collected by different research teams. These were used to develop and test the proposed method (see Table \ref{tab1}). Among them, Training-A is the only database which contains simultaneously recorded PCGs and ECGs. 
The other 11 sub data sets only contains PCGs, namely Training-B\textasciitilde E* and F, and Test-B\textasciitilde E*, G and I.

The state labels of data sets were assigned as onsets of S1, systole, S2 and diastole.
S1 occurs immediately after R-peaks (ventricular depolarization) of the ECG, while S2 occurs at approximately at the end-T-waves of the ECG (the end of ventricular depolarization) \cite{lee2019electrocardiogram, springer2015logistic}.
Therefore, for PCGs with synchronous ECGs, the automated-detected R peaks and end-T-waves were the basis of annotations of S1 and S2 onsets. 
The segmented results  solved by LR-HSMM were utilized as automatic marks for annotation reference in recordings containing only PCGs.
The incorrect annotations happen in Training-A when R peaks and end-T-wave positions of an abnormal ECG period were misdetected. For data sets besides Training-A, although the annotations provided for the challenge were manually corrected by the organizers, some of them were still questionable and required re-annotations.
These annotations were hand-corrected by visual and audible inspection of PCG waveforms.
The re-annotation instances in Training-A are illustrated with reference ECG in Fig.2.

\begin{table}[!ht]
    \centering
    \begin{threeparttable}[b]
    \renewcommand{\arraystretch}{1.25}
    \setlength{\tabcolsep}{0.5mm}
    \caption{Statistics of re-annotated databases after excluding unsure recordings.}
    \label{tab1}
    \begin{tabular}{cccccc}
        \hline
         Database & Total Recordings & Total Beats & Recordings Used & Beats Used \\
        \hline
         Training-A & 392 & 14,560 & 392 & 14,560 \\
         Training-B & 368 & 3,353 & 368 & 3,353 \\
         Training-C & 27 & 1,801 & 27 & 1,801 \\
         Training-D & 52 & 853 & 52 & 853 \\
         Training-E$^*$ & 1,927 & 59,637 & 500 & 15,256 \\
         Training-F & 108 & 4,452 & 108 & 4,452 \\
         Test-B     & 206 & 1,247 & 206 & 1,247 \\
         Test-C     & 15 & 1,007 & 15 & 1,007 \\
         Test-D     & 24 & 268 & 24  & 268 \\
         Test-E$^*$ & 882 & 25,261 & 200 & 6,060 \\
         Test-G     & 174 & 2,116 & 174 & 2,116 \\
         Test-I     & 33 & 1,207 & 33  & 1,207 \\
        \hline
         Total & 4,208 & 115,762 & 2,099 & 52,180 \\
        \hline
    \end{tabular}
    \begin{tablenotes}
        \item[1] \(*\) in Training-E$^*$ and Test-E$^*$ indicates that part of original Training-E and Test-E were utilized in this work.
    \end{tablenotes}
    \end{threeparttable}
\end{table}
 
Since the majority of heart sounds in Training-E (N=4,074) and Test-E (N=901) are normal, a total of 500 and 200 recordings were randomly extracted from Training-E and Test-E respectively to alleviate the re-annotation work while ensuring the accuracy of the evaluation.

\subsection{Derivative Date Set}
In order to further excavate characteristics of each data set, several indicators were designed as follows:
\begin{align}
    \label{eqn1} &W = \frac{1}{N}\sum_{n=1}^N{y^2[n]}, \\
    \label{eqn2} &F_\text{S2} = \frac{W_{S2}}{W_{diastole}},
\end{align}
\begin{align}
    \label{eqn3} &D_\text{noise\&murmur} = \frac{W_\text{systole}+W_\text{diastole}}{W_\text{S1}+W_\text{S2}}, \\
    \label{eqn4} &D_\text{rhythm} = \sqrt{\frac{1}{M-1}\sum_{i=1}^{M-1}{(ss[i]-E\left\{ss\right\})^2}}, \\
    \label{eqn5} &D_\text{rate} = \biggl\| \mathop{\max}(1.2, E\left\{ss\right\})+\mathop{\min}(0.6, E\left\{ss\right\}) - 1.8 \biggr\|,
\end{align}
where \(W\) is the average power of the heart sound \(y[n]\) with \(N\) points and \(M\) heart beats.
\(F_\text{S2}\) defines S2 sound articulation as the ratio of S2 sound's energy to diastolic energy.
\(D_\text{noise\&murmur}\) is the ratio of power between systole/diastole and S1/S2, which represents the index of murmur severity and signal quality.
\(D_\text{rhythm}\) is estimated by the summation of the standard deviation of S1 onset intervals (\(ss\)) to measure the severity of arrhythmia.
\(D_\text{rate}\) is relative difference of \(E\left\{ss\right\}\) from normal range, which reflects the degree of abnormality in heart rate (bradycardia and tachycardia). 
In Eqn. \ref{eqn5}, 1.2s and 0.6s are average S1 onset intervals of 50 and 100 heart beats per minute, respectively.

According to the statistical results of the indicators, for normal heart sounds, \(D_\text{noise\&murmur}\) is below 0.3 and \(F_\text{S2}\) is larger than 2.0.
And for the cases contaminated by sever murmurs or noise, \(D_\text{noise\&murmur}\) is always over 0.8.
Therefore, the noise\&murmur level can be divided into low (\(D_\text{noise\&murmur}\leq 0.8\)) and high (\(D_\text{noise\&murmur}>0.8\)).
During arrhythmia, \(D_\text{rhythm}\) is over 0.12, which is also the indicated value for PP interval deviation of arrhythmic ECG.
\(D_\text{rate}\) would be greater than 0 when the heart rate is abnormal.

\begin{figure}[!t]
    \centering
    \includegraphics[width=2.5in]{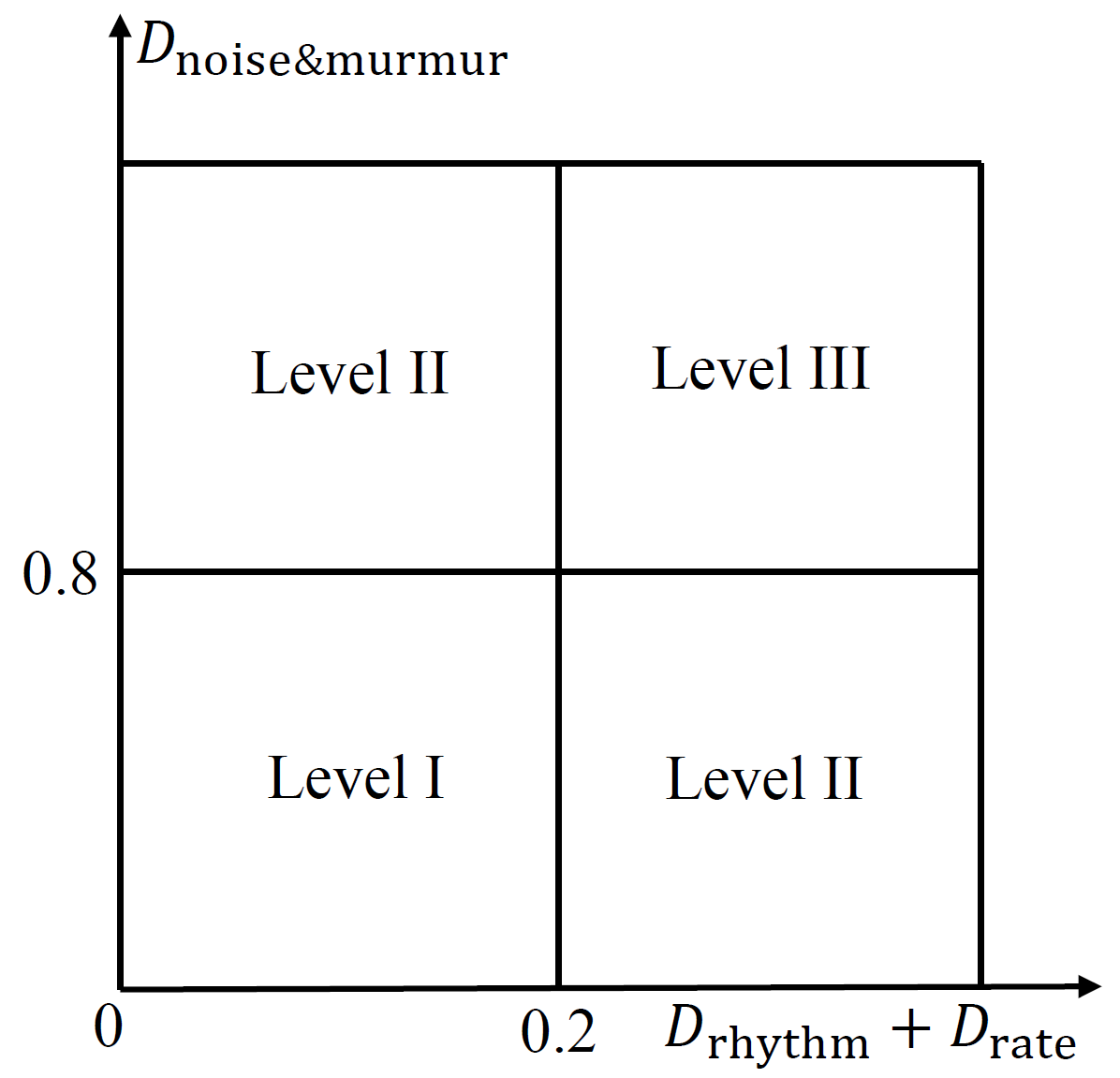}
    \caption{The partition rules of the three difficulty levels (LEVEL-I, LEVEL-II and LEVEL-III) based on the extent of noise/murmur and evaluation of heart rhythm/rate in heart sounds.}
    \label{fig3}
\end{figure}

Three derivative data sets were constructed according to the designed indicators. They were named LEVEL-I, LEVEL-II and LEVEL-III, corresponding to easy, medium and difficult in terms of both automatic and manual heart sound segmentation. 
The threshold of \(D_\text{noise\&murmur}\) is set to 0.8 to distinguish heart sounds with complicated severe noise and murmur. 
\(D_\text{rhythm}+D_\text{rate}\) is an indicator of abnormal heart rhythm and heart rate.
Thus, the threshold of \(D_\text{rhythm}+D_\text{rate}\) is assigned a value of 0.2s. Fig.\ref{fig3} provides a graphical illustration of the above partition rules.

\begin{figure*}[!ht]
  \centering
  \begin{minipage}[b]{6.8in}
    \centering
    \includegraphics[width=6.8in]{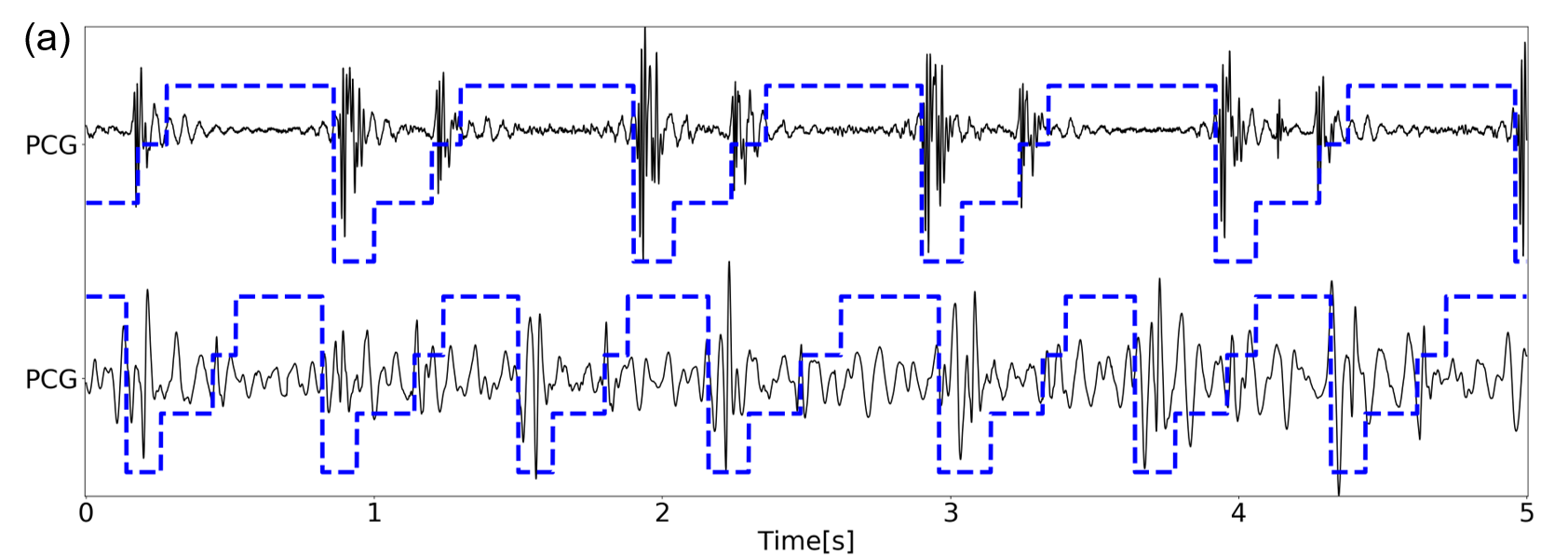} \\
    \includegraphics[width=6.8in]{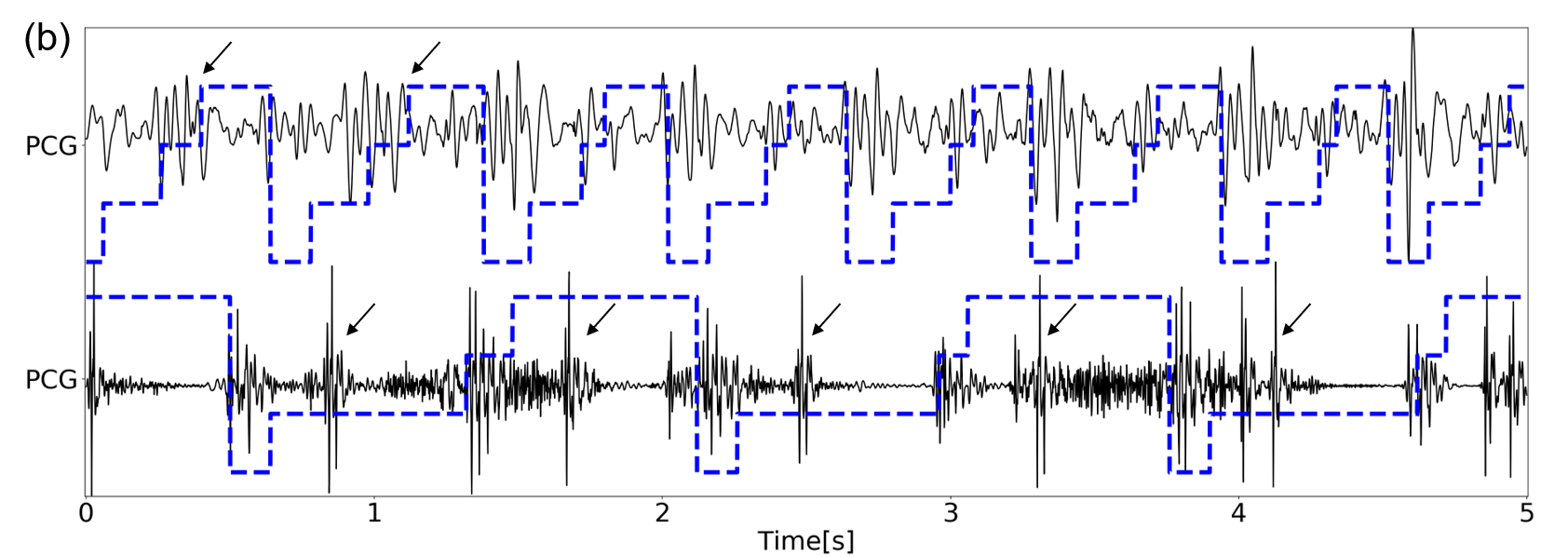} \\
    \includegraphics[width=6.8in]{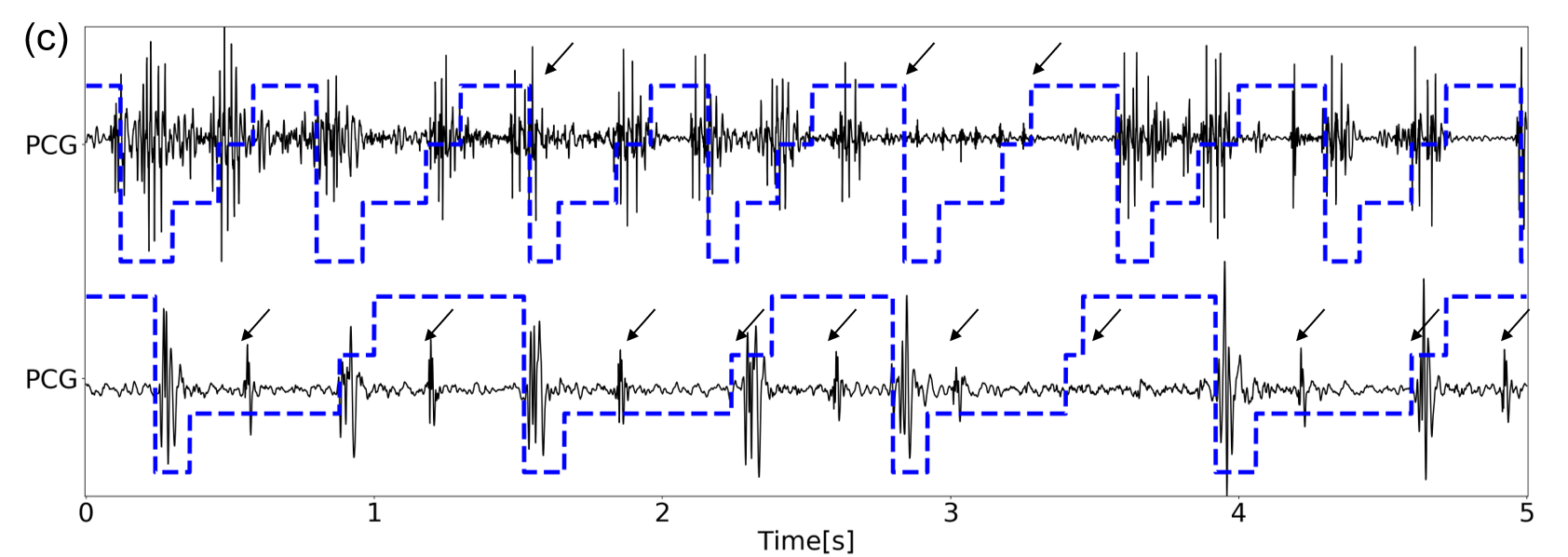}
  \end{minipage}
  \caption{Sub-figures (a), (b) and (c) correspond to three instances from LEVEL-I, LEVEL-II and LEVEL-III respectively. The blue dashed line indicates the states assigned by the LR-HSMM method and the arrows identify the unsuccessful segmentation. Note that the increased difficulty significantly impacts the performance of the LR-HSMM method. }
  \label{fig4}
\end{figure*}

All of the heart sound recordings chosen for different levels were segmented into multiple 10 second files. The resultant numbers of recordings and beats are summarized in Table \ref{tab2}. The specific instances in the three difficulty levels are displayed in Fig.\ref{fig4}.

\begin{table}[!ht]
    \centering
    \renewcommand{\arraystretch}{1.25}
    \setlength{\tabcolsep}{8mm}
    \caption{Summary of databases corresponding to three difficulty levels.}
    \label{tab2}
    \begin{tabular}{ccc}
        \hline
         Database & Recordings & Beats \\
        \hline
         LEVEL-I  & 200 & 2,296 \\
         LEVEL-II & 200 & 2,459 \\
         LEVEL-III & 150 & 1,906 \\
        \hline
         Total & 550 & 6,661 \\
        \hline
    \end{tabular}
\end{table}

We counted the proportion of various anomalies in each data set based on the indicators, including heart sounds with high-level noise\&murmur, arrhythmic heart sounds, heart sounds with abnormal heart rate and heart sounds with vague S2.
It was found that Training-B and Test-B have the lowest signal quality among the data sets and all of the data sets contain a certain amount of heart sounds with arrhythmia and abnormal heart rate expect for Training-A. The specific data is shown in Table \ref{tab3}. 

\begin{table*}[!ht]
    \centering
    \begin{threeparttable}[b]
    \renewcommand{\arraystretch}{1.25}
    \setlength{\tabcolsep}{4mm}
    \caption{Summary of the characteristics in each data set. The number of recordings and proportion is reported. The characteristics include noise\&murmur, arrhythmia, abnormal heart rate(Abn. HR) and vague S2.}
    \label{tab3}
    \begin{tabular}{cccccccccc}
        \hline
         \multicolumn{1}{c}{Database} & \multicolumn{1}{c}{Recordings} & \multicolumn{2}{c}{High-level Noise\&Murmur} & \multicolumn{2}{c}{Arrhythmia} & \multicolumn{2}{c}{Abn. HR} & \multicolumn{2}{c}{Vague S2} \\ 
         \cline{3-10}
         &  & Count & Prop.(\%) & Count & Prop.(\%) & Count & Prop.(\%) & Count & Prop.(\%) \\ 
         \hline
         Training-A & 392 & 3 & 0.77 & 25 & 6.38 & 21 & 5.36 & 32 & 8.16 \\
         Training-B & 368 & 184 & 50.0 & 4 & 1.09 & 30 & 8.15 & 274 & 74.47\\
         Training-C & 27 & 2 & 7.41 & 4 & 14.85 & 2 & 7.41 & 6 & 22.22 \\
         Training-D & 52 & 1 & 1.92 & 8 & 15.38 & 11 & 21.15 & 5 & 9.62 \\
         Training-E$^*$ & 500 & 24 & 4.80 & 13 & 2.60 & 95 & 19.00 & 51 & 10.20 \\
         Training-F & 108 & 1 & 0.93 & 22 & 20.37 & 11 & 10.19 & 0 & 0.00 \\
         Test-B & 206 & 72 & 34.95 & 6 & 2.91 & 22 & 10.68 & 118 & 57.28 \\
         Test-C & 15 & 2 & 13.33 & 2 & 13.33 & 1 & 6.67 & 2 & 13.33 \\
         Test-D & 24 & 0 & 0.00 & 3 & 12.50 & 3 & 12.50 & 1 & 4.17 \\
         Test-E$^*$ & 200 & 1 & 6.67 & 5 & 2.50 & 39 & 19.50 & 25 & 12.50 \\
         Test-G & 174 & 3 & 1.72 & 11 & 6.32 & 11 & 6.32 & 23 & 13.22 \\
         Test-I & 33 & 6 & 18.18 & 0 & 0.00 & 1 & 3.03 & 14 & 42.42 \\
         Level-I & 200 & 0 & 0.00 & 0 & 0.00 & 13 & 6.50 & 23 & 11.50 \\
         Level-II & 200 & 33 & 16.50 & 27 & 13.50 & 14 & 7.00 & 49 & 24.50 \\
         Level-III & 150 & 105 & 70.00 & 45 & 30.00 & 14 & 9.33 & 19 & 12.67 \\
         \hline
    \end{tabular}
    \begin{tablenotes}
        \item[1] \(*\) in Training-E$^*$ and Test-E$^*$ indicates that part of original Training-E and Test-E were utilized in this work.
    \end{tablenotes}
    \end{threeparttable}
\end{table*}

\section{Methods}
The proposed method involves two main parts: a signal pre-processing routine, and the TFAN segmentation. The signal pre-processing employed three filters: an adaptive Wiener filter, a bandpass filter and a wavelet filter. The TFAN is an original network designed for heart sound segmentation with an encoder-decoder architecture. In order to learn the state transition information in PCG, the loss function of the TFAN was carefully designed. 

\subsection{Signal Pre-processing}
\begin{figure}[!t]
  \centering
  \begin{minipage}[b]{0.5\textwidth}
    \includegraphics[width=3.5in]{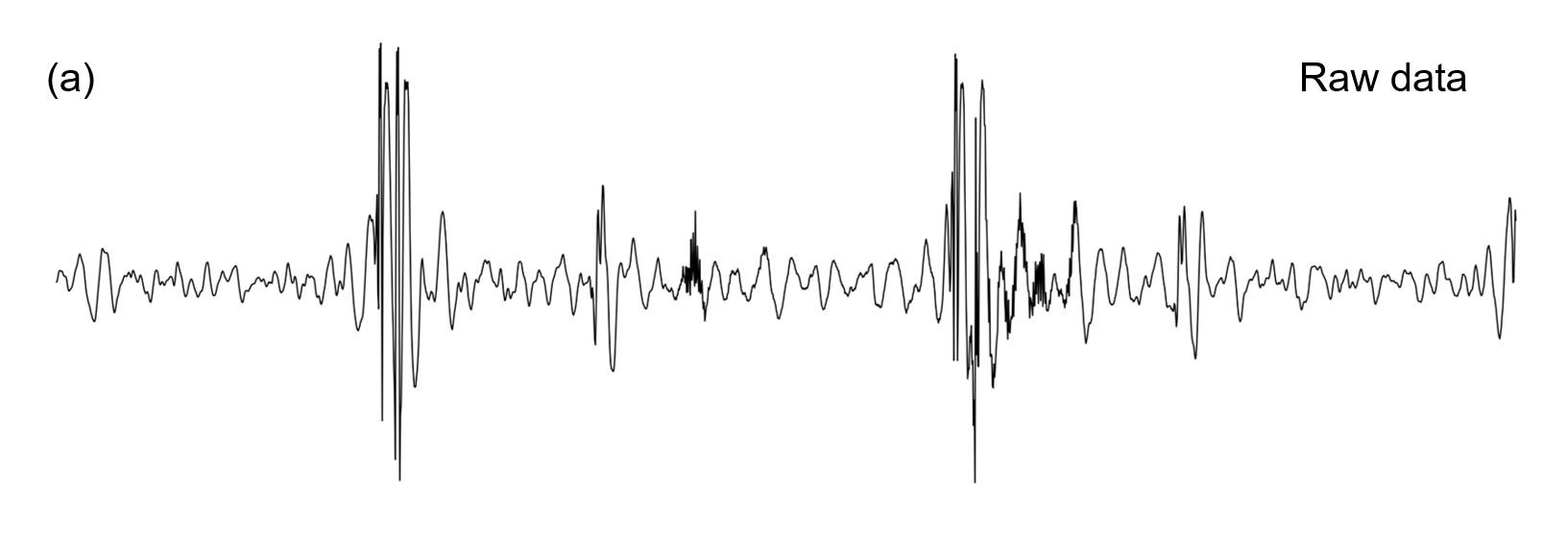} \\
    \includegraphics[width=3.5in]{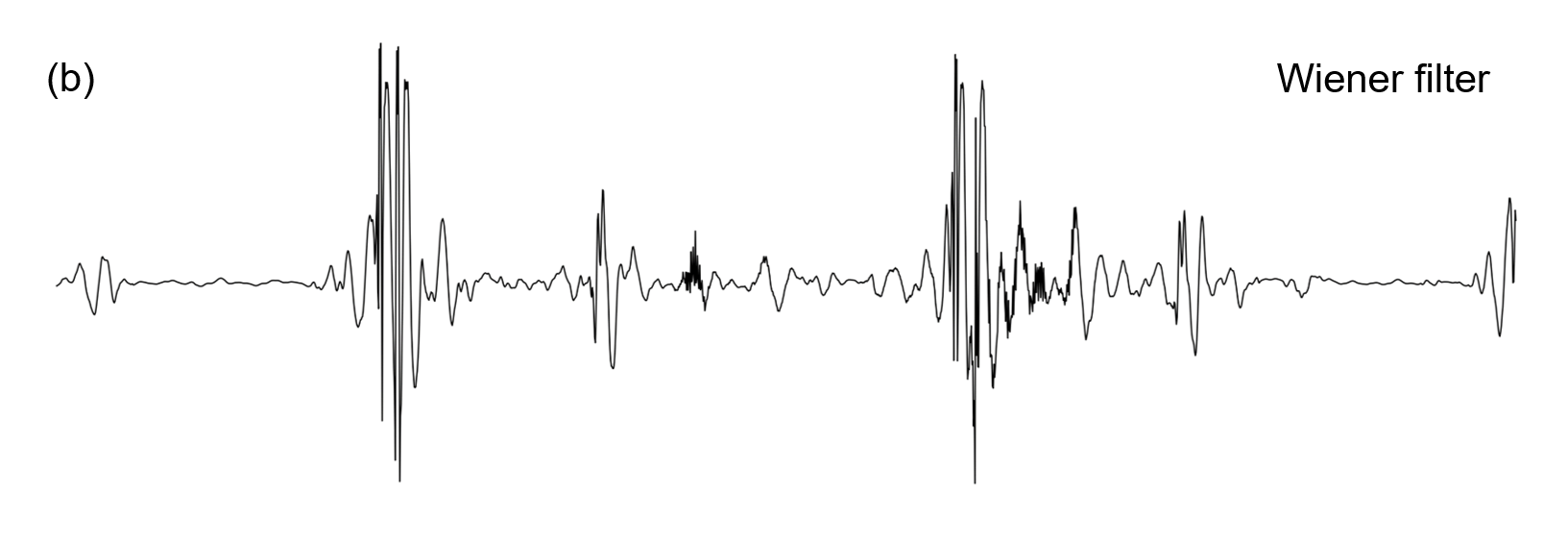} \\
    \includegraphics[width=3.5in]{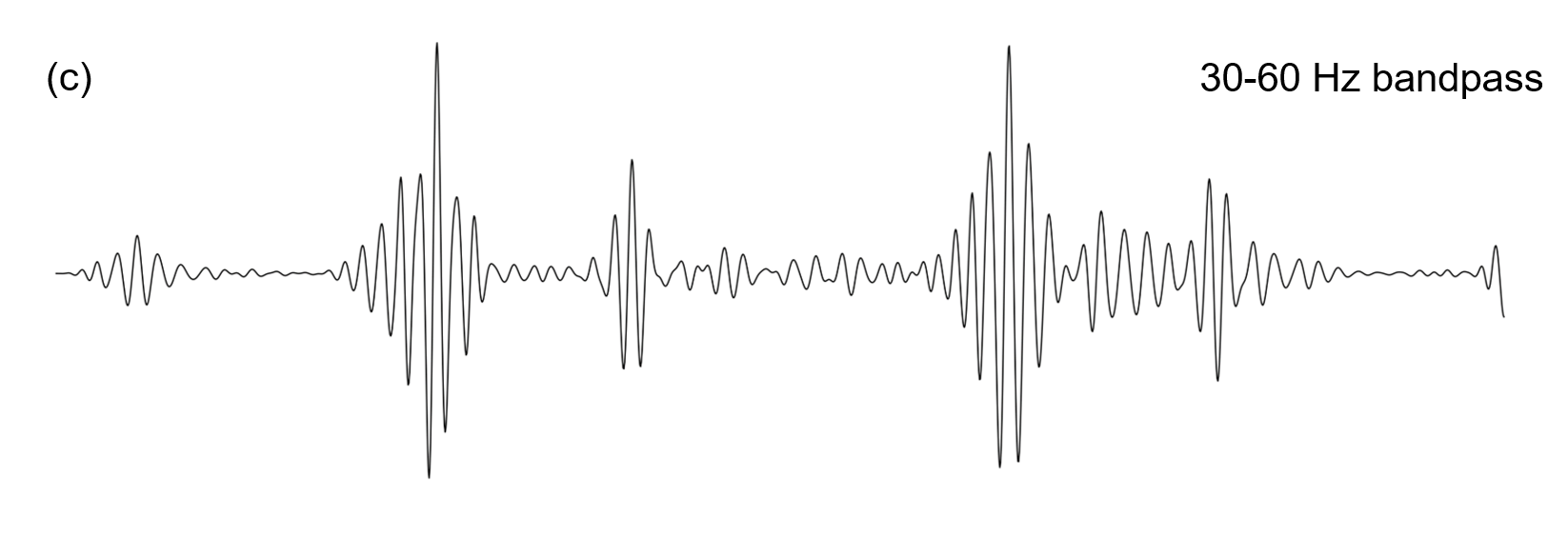} \\
    \includegraphics[width=3.5in]{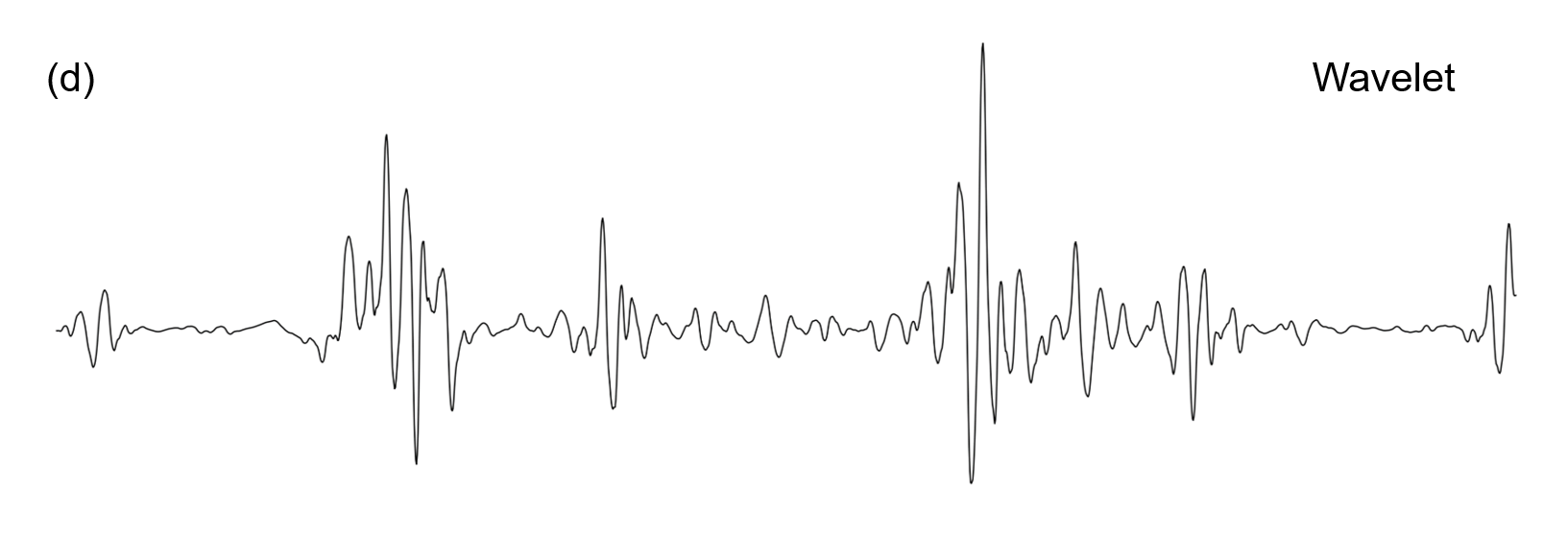}
  \end{minipage}
  \caption{Sub-figures (a), (b), (c) and (d) demonstrate the examples of raw PCG signal, the signal processed by an adaptive Wiener filter, the signal processed by a 30-60 Hz bandpass filter and the signal processed by a reverse biorthogonal wavelet filter. The output of each filter constitute the three channels of input data for temporal framing network.}
  \label{fig5}
\end{figure}

The segmentation algorithm used a combination of three filtered PCG signals as inputs.
The three filters included an adaptive Wiener filter, a bandpass filter and a wavelet filter. 
As shown by the instances reported in Fig. \ref{fig5}, the adaptive Wiener filter was designed to suppress the in-band noise, especially reduce the impact of tail sounds in systole and diastole. This approach increased the amplitude resolution of alternate segments between heart sound states.
The bandpass filter and the wavelet filter were applied to enhance S1 and S2 sounds and provided complementary information of their waveform features.

\subsubsection{Adaptive Wiener Filter}
Consider a zero-mean clean heart sound signal \(x(n)\) contaminated by noise \(v(n)\) (uncorrelated with \(x(n)\)), so that the noisy heart sound \(y(n)\) at the discrete time \(n\) is
\begin{equation}
  \label{eqn6}
  y(n)=x(n)+v(n), n=0,...,N-1,
\end{equation}

The estimation of the error signal \(e_x(n)\) between the clean heart sound at the discrete time \(n\) is given by
\begin{equation}
  \label{eqn7}
  e_x(n)=x(n)-\hat{x}(n)=x(n)-\boldsymbol{h^Ty}(n),
\end{equation}
where superscript \(^T\) denotes transpose of a vector or a matrix,
\begin{equation*}
    \boldsymbol{h}=[h_0, h_1,...,h_(L-1)]^T
\end{equation*}
is an finite impulse response (FIR) filter of length L, and
\begin{equation*}
    \boldsymbol{y(n)}=[y(L-1), y(L-2),...,y(0)]^T
\end{equation*}
is a vector of window from observation signal y(n) containing L samples.

Assuming the optimal estimate of the clean heart sound \(x(n)\) is \(\hat{x}_o(n)\), the optimal filter \(\boldsymbol{h_o}\) for \(\hat{x}_o(n)\) is the Wiener filter which is obtained by
\begin{equation}
  \label{eqn8}
  \boldsymbol{h_o}=\mathop{\arg\min_{\boldsymbol{h}}}E\lbrace{e_x^2(n)}\rbrace.
\end{equation}

According to Wiener-Hopf equation, we have
\begin{equation}
  \label{eqn9}
  \boldsymbol{R_yh_o}=E\left\{y(n)x(n)\right\}=\boldsymbol{r_y}-\boldsymbol{r_v},
\end{equation}
where \(\boldsymbol{R_y}\) is the correlation matrix of the observed signal \(y(n)\). \(\boldsymbol{r_y}\) and \(\boldsymbol{r_v}\) are the correlation vectors, which are also the first columns of \(\boldsymbol{R_y}\) and the correlation matrix of the noise \(\boldsymbol{R_v}\) respectively.

Now \(\boldsymbol{h_o}\) can be inferred as
\begin{equation}
  \label{eqn10}
  \boldsymbol{h_o}=\boldsymbol{u_1}-\boldsymbol{R_y^{-1}r_v},
\end{equation}
where \(\boldsymbol{u_1}=[1,0,...,0]^T\).

Assuming that the additive noise is white over a very short time duration in comparison to the heart sounds, we have
\begin{equation}
\label{eqn11}
  \boldsymbol{r_v}=\sigma_v^2\boldsymbol{u_1},
\end{equation}
and
\begin{equation}
\label{eqn12}
  \begin{split}
  \boldsymbol{h_o}&=\boldsymbol{u_1}-\sigma_v^2\boldsymbol{R_y^{-1}u_1}\\
  &=[1-\frac{\sigma_v^2}{\boldsymbol{R_y}[0]}, 1-\frac{\sigma_v^2}{\boldsymbol{R_y}[1]},...,1-\frac{\sigma_v^2}{\boldsymbol{R_y}[L-1]}],
  \end{split}
\end{equation}
where $\sigma_v^2=E\{v^2(n)\}$.

Because the noise \(v(n)\) is not directly observable, \(\sigma_v^2\) is ideally calculated while there is no heart sound signal. In order to avoid being disturbed by sudden changes in the recording, the local window is segmented in fixed length and \(\sigma_v^2\) is estimated as the lower quartile of the local variances \(Q1(\boldsymbol{lvar})\) for all segments. Finally the estimated heart sound of the local window \(\hat{x}(n)\) is given by
\begin{equation}
\label{eqn13}
  \begin{split}
  \hat{x}(n)&=\boldsymbol{h_o}(n)^T\boldsymbol{y}(n)\\
  &=(1-\frac{Q1(\boldsymbol{lvar})}{\boldsymbol{R_y}[n]})\boldsymbol{y}(n), n=0,1,...,L.
\end{split}
\end{equation}

\subsubsection{Bandpass Filter}
The majority of the frequency content in S1 and S2 sounds is below 150 Hz, usually with a peak around 50 Hz \cite{arnott1984spectral}. Thus, a Bandpass filter was applied to create a signal with 30-60 Hz pass-band, to be used as one input channel to provide the potential optimal positions of S1 and S2.

\subsubsection{Wavelet Filter}
The first step in the wavelet filter for heart sounds is a discrete time wavelet transform (DWT). 
Following Springer {\it et al.} \cite{springer2015logistic}, the reverse biorthogonal wavelet with three vanishing moments for the decomposition (analysis) wavelet and nine vanishing moments for the reconstruction (synthesis) wavelet
(\('rbio3.9'\)) was used.
In order to remove the insignificant noise, the detail coefficients below an adaptive threshold at some scales were set to zero. 
The threshold was set to be the median energy, which was estimated by averaging the absolute coefficients at different scales. 
The final filtered heart sounds were reconstructed by the inverse DWT.

\subsection{Temporal-Framing Adaptive Network}
\subsubsection{Model Architecture}
\begin{figure*}[!t]
  \centering
  \includegraphics[width=1.0\textwidth]{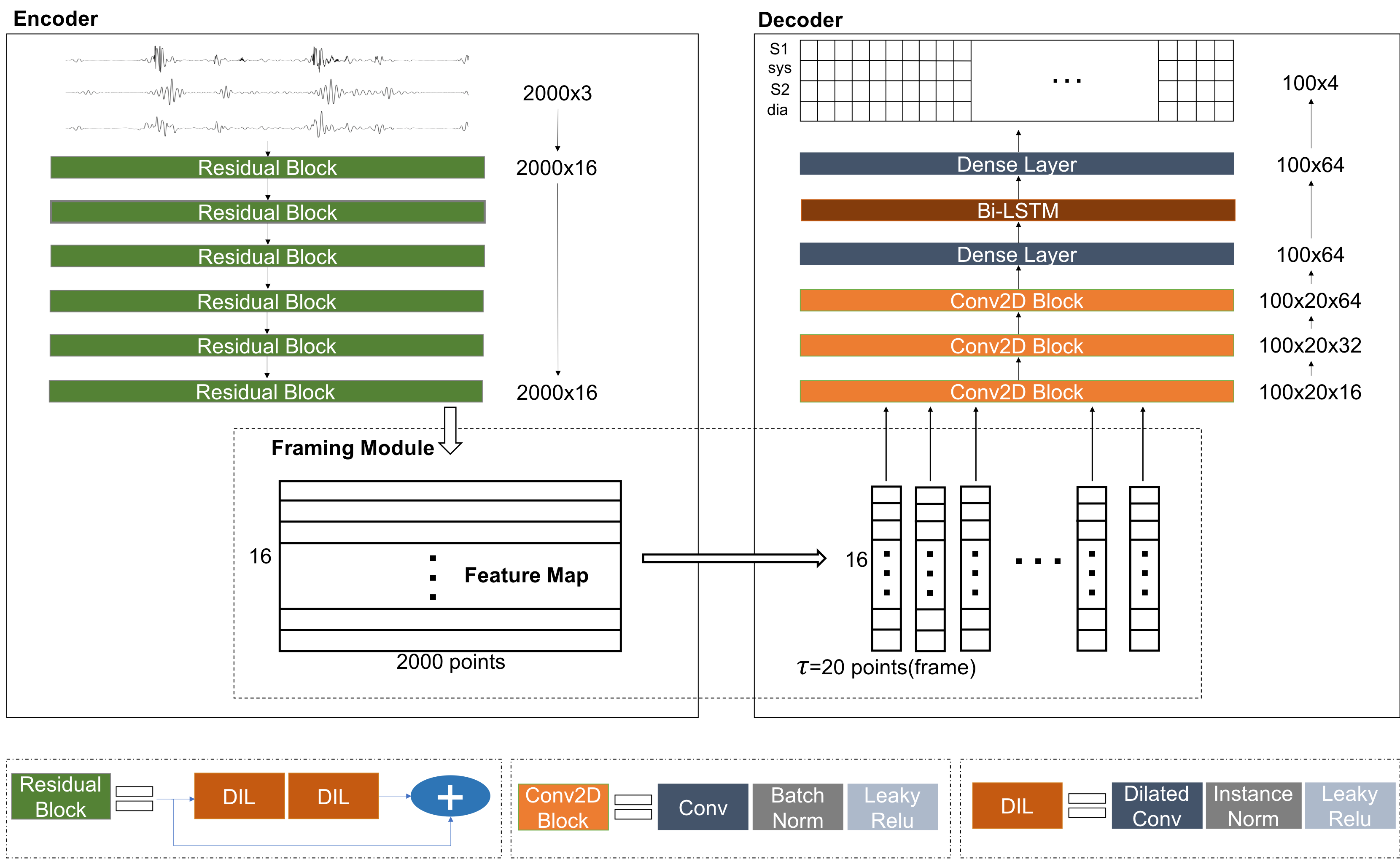}
  \caption{The architecture of the proposed TFAN. The PCG signal is framed into multi-frames after feature mapping in the Encoder. Then the whole frames in one batch generate a new batch of features to be input into the decoder. The final outputs are the predicted logits of four states in each frame.}
  \label{fig6}
\end{figure*}

The TFAN was designed with an encoder-decoder architecture. The encoder (Fig. \ref{fig6}) is a transformer module for the purpose of mapping the original signals into a feature space. The decoder (Fig. \ref{fig6}) is designed to segment the output feature mapped by the encoder into four states (S1, systole, S2, diastole). Within the network, a framing module is deployed between the encoder and the decoder.

Before loaded to the TFAN model, each processed heart sound recording was sliced into segments of two seconds and resampled to 1,000 Hz. The input data is denoted as \(x\left(n\right)\) for \(n=0,...,1,999\) and \(x(n)\in{\mathbb{R}^3}\).

\subsubsection{Encoder}
A residual convolution block is used as a basic unit for feature mapping (Fig.\ref{fig6}). The residual block \cite{he2016deep} contains a branch leading out to a series of transformations \(F\), whose outputs are added to the input \(x\) of the block, so the original mapping is recast into \(x+F(x)\).
This effectively allows layers to learn modifications to the identity mapping, rather than the entire transformation, which is more advantageous for identifying similar states in the heart sound (\emph{e.g.}, S1 and S2).

In the TFAN, instance normalization (IN) and dilated convolution were utilized in each residual block. The reasons for using IN are: 1) The segmentation model is trained with limited batch size and IN normalizes across each training sample instead of the mini-batch, therefore biased estimations of mean and variance of mapped features are avoided; 2) IN normalizes across each channel, so the independence of each channel is maintained. For the input data \(x\in{R}^{N\times{T}\times{C}}\), IN calculates the mean and variance across the time dimension of each sample and retains the dimensions of the batch \(N\) and channel \(C\) as
\begin{align}
  \mu_{nc}(x) & =\frac{1}{T}\sum_{t=1}^T{x_nct},\\
  \sigma_{nc}(s) & =\sqrt{\frac{1}{T}\sum_{t=1}^T{(x_{nct}-\mu_{nc}(x))^2}+\varepsilon},
\end{align}
where \(\varepsilon\) is the biased value to avoid division by \(0\) when normalizing the weights.

\begin{figure}[ht]
  \centering
  \includegraphics[width=3.5in]{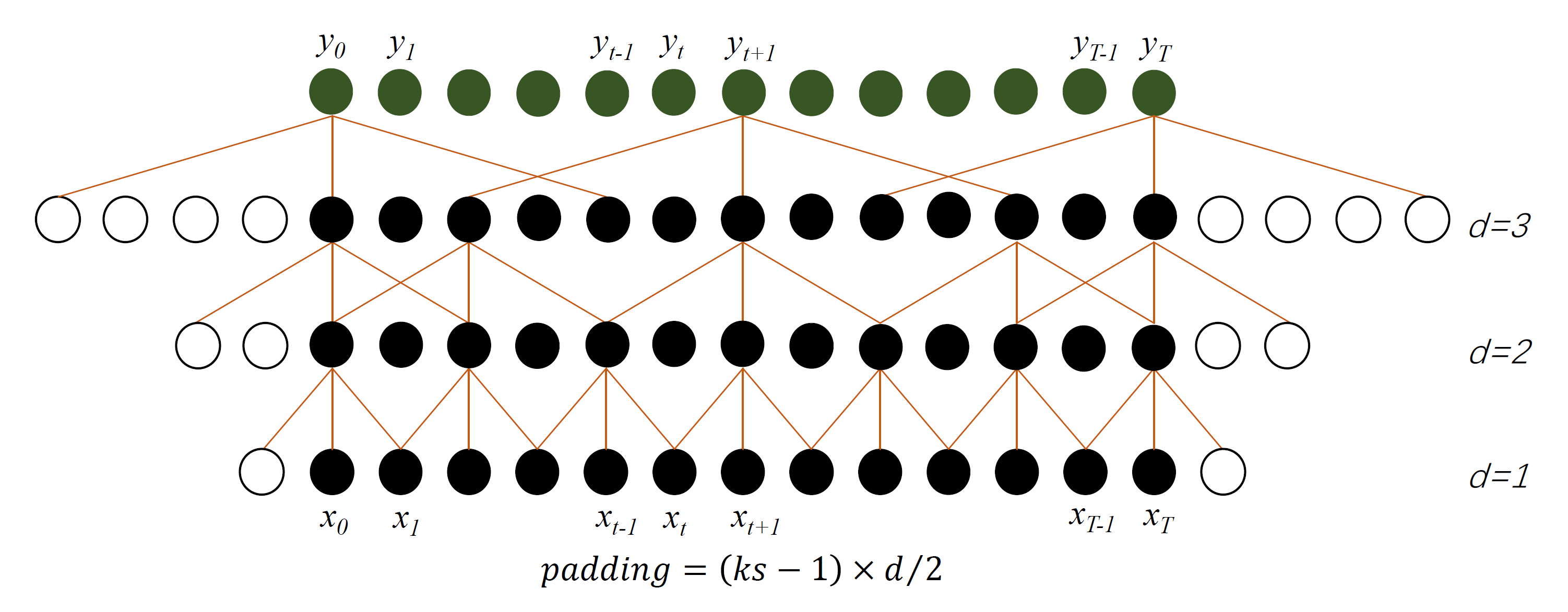}
  \caption{The padding strategy of the convolution operation in temporal residual block. The purpose using the above padding pattern is to maintain the dimension while feature mapping. }
  \label{fig7}
\end{figure}

Dilated convolution can enlarge the receptive field of convolution layers and preserve the size of feature maps without loss of resolution. This is critical for the subsequent framing module and decoder. Meanwhile, bidirectional padding is chosen as the padding strategy in dilated convolution. The padding length is decided by the convolutional kernel size and dilation rate. Fig. \ref{fig7} illustrates the padding method for different dilation rates (\(d\)) in the case of a convolution kernel size (\(ks\)) of 3.

\subsubsection{Decoder}
Before passed through the Decoder, the feature map of the heart sound produced by Encoder is framed by a fixed length \(\tau\). Then the frame-level features can be further mapped by 2D convolution blocks (Fig. \ref{fig6}) of decoder. The output is then fed to a bidirectional long short-term memory (Bi-LSTM) layer to learn sequential characteristics of the frame-level features.

Assuming the mappings of the encoder and decoder are denoted as \(f=\mathbb{F}(x(n))\) and \(g=\mathbb{G}(f)\), the feature map is transformed as below after the frame-level decoder:
\begin{align}
  f & \rightarrow{[f_0, f_1,...,f_{\frac{N}{\tau}}]},\\
  \hat{s}(m) & =[g(f_0), g(f_1),...,g(f_{\frac{N}{\tau}})], m=0,...,\frac{N}{\tau}.
\end{align}
\(\hat{s}(m)\) is defined as the sequence of the logits output from the model, where \(s(m)\) is the ground truth of the heart sound states.

\subsubsection{Loss Function}
According to the periodic nature of heart sounds, the identification of state for each frame is determined not only based on the features but the state transition information between the current and preceding frames. Therefore, the loss in the TFAN is the combination of the classification loss and the state transition loss between the current frame and the previous frame: 
\begin{equation}
\centering
\label{eqn18}
\begin{aligned}
  L(y, \hat{y})=-\frac{1}{T}\frac{1}{N}\sum_{\tau=1}^T\sum_{i=1}^N \biggl\{ {C_{1}\times{y_{i\tau}\log{\hat{y}_{i\tau}}}}\\
  {+C_{2}\times\frac{y_{i\tau}+y_{i(\tau-1)}}{2}\log\frac{\hat{y}_{i\tau}+\hat{y}_{i(\tau-1)}}{2}} \biggr\}, \\
\end{aligned}
\end{equation}
where \(y\) and \(\hat{y}\) represent the annotated state and the predicted mask of the frame, respectively. \(T\) and \(N\) are the number of frames and the number of heart sound states, separately, which \(T=100\) and \(N=4\) in the TFAN-based method. \(y_{i0}\) and \(\hat{y}_{i0}\) are padded as the ground truth and the predicted logit of the first frame. The weighting parameters \(C_{1}\) and \(C_{2}\) could help adjust the constraint degree of the state transition information and the features of each frame in state prediction. In the TFAN-based method, \(C_{1}=1\) and \(C_{2}=2\).

\subsubsection{Dynamic Inference}
Since the length of the input data of our model is fixed, the heart sound recording needs to be divided into segments for dynamic inference. In order to minimize the impact on segmentation of data around slicing boundary, 50\% overlapping windows are adopted.
For the overlapping windows, the logits of different states are simply averaged. If the length of the remaining recorded data is less than the 50\% overlapping duration, the input segment of fixed length is taken before the last point. 
Meanwhile, the logits of the remaining data are retained and concatenated with the previous results so that all points of the recording can be detected.

Knowning \(s(t)\in{\left\{0, 1, 2, 3\right\}}\), each element in \(s(t)\) corresponds to S1, systole, S2, diastole respectively. 
The labels are then one-hot encoded. 
The outputs from TFAN are the logits of four states for each frame. Assuming the input data is \(F\) and the length of state sequence after framing is \(M\), the inference step needs to find out \(s(m)\) by \( \mathop{\arg\max}_{s}P(s_1,s_2,...,s_M|F) \). Since the total search of \(s_{1\sim{M}}\) for the best state sequence required \(4^M\) times, the search time complexity would be high when \(M\) is large. The Viterbi algorithm is therefore adopted to shorten the solving time.
Based on the Viterbi algorithm, the inference method can be transformed to
\begin{equation}
  \centering
  \mathop{\max}P(s_1,...,s_M|F)=\mathop{\max}\left\{q(v,M)|v\right\},
\end{equation}
where
\begin{equation}
  \centering
  q(v=j,m)=\mathop{\max}\left\{q(v=i,m-1)\times{a(i,j,m)}|i\right\},
\end{equation}
for \(v=0,1,2,3\) and \(i=0,1,2,3\). Note that \(q(v,M)\) represents the maximum probability for the state sequence ending with \(v\), and \(a(i,j,m)\) defines the transforming probability from state \(i\) at step \(m-1\) to state \(j\) at step \(m\). For heart sound states, the state transition probability matrix is given by
\begin{equation}
  \centering
  A=\begin{bmatrix}
      a_{11} & a_{12} & \cdots & a_{14} \\
      a_{21} & a_{22} & \cdots & a_{24} \\
      \vdots & \vdots & \vdots & \vdots \\
      a_{41} & a_{42} & a_{43} & a_{44}
    \end{bmatrix}=
  \begin{bmatrix}
    0.5 & 0.5 & 0 & 0\\
    0 & 0.5 & 0.5 & 0\\
    0 & 0 & 0.5 & 0.5\\
    0.5 & 0 & 0 & 0.5
  \end{bmatrix}.
\end{equation}

The predicted state sequence \(s\) is inferred by the following function
\begin{equation}
  \centering
  s_m=\mathop{\arg\max}_{v}q(v,m).
\end{equation}

\section{Experiments}
The proposed segmentation methods were compared with two methods appeared in the literature. Namely, the BiGRNN-based method using spectrogram and envelop features described in \cite{messner2018heart} and the LR-HSMM method, which is currently considered as the state-of-art PCG segmentation method.
For fairly comparing the performance of TFAN and BiGRNN, the proposed dynamic inference approach was conducted in both methods.

Besides, as a generative model, LR-HSMM is essentially trained to explicitly model the probability distribution of each heart sound state given four extracted features on each time step. 
Therefore, the LR-HSMM method is not sensitive to the size of training set. 
In the previous study, LR-HSMM is trained by recordings from only 60 patients \cite{springer2015logistic}. 
Instead, for DL-based methods, all-round data sets are generally required for precisely learning data distribution.
Such as \cite{messner2018heart}, the training set for BiGRNN consisted of recordings randomly selected from Training-B\textasciitilde F of the challenge.
In order to improve the extensiveness of DL-based methods, we introduced the framing module in the proposed TFAN, converting estimation of state sequence to estimation of state on each time step.
For finding out the influence of various feature learning approaches on segmentation and the capacity of the proposed method in few-shot learning, we limited the number of training recordings.

The experiments comparing the performances of the three methods were conducted in two scenarios. 
The first scenario was to test the methods on Training-A\(^*\) and other independent sub data sets from 2016 PhysioNet/CinC Challenge. 
The second scenario was to test on the data sets of three difficulty levels (LEVEL-I, LEVEL-II and LEVEL-III).

\subsection{Training Setup}
Since the gold-standard reference positions of onsets of S1 and S2 sounds were derived from the synchronous ECGs \cite{springer2015logistic}, to ensure the preciseness of the experiment, the training set was consisted by heart sound recordings with synchronous ECGs splitted from Training-A.
The ultimate size of training set was restricted to 50 recordings for all the methods, and the remaining recordings were utilized as Training-A* for testing.


Five-fold cross-validation was adopted as the training approach. 
For ensuring the recording used for validation is not used to train, 50 heart sound recordings were split into 40 recordings for training and 10 recordings for validation in each fold at first. 
Then the heart sounds in training and validation set were pre-processed and sliced into 2s segments for neural network training.
After five-fold cross-validation training, the model with best performance on the validation fold was chosen.

The loss functions of BiGRNN and TFAN were unified into the proposed one.
Weights of both models were updated by the Nesterov Momentum optimizer with factor of 0.9 and learning rate of 0.001.
In order to avoid overfitting, the following early stop strategy was adopted.
When the model failed to achieve the best validation accuracy in 20 consecutive epochs, the training is terminated.

\subsection{Evaluation Metrics}
To evaluate the performance of the TFAN-based method against the LR-HSMM method, three measurements are considered, which are defined as:
\begin{eqnarray}
  SE = \frac{TP}{TP+FN} \times{100\%},\\
  P_+ = \frac{TP}{TP+FP} \times{100\%},\\
  F_1  = \frac{2\times{SE}\times{P_+}}{SE+P_+} \times{100\%},
\end{eqnarray}
where \(TP\) (true positive), \(FP\) (false positive) and \(FN\) (false negative) are determined by the following rules\cite{liu2017performance}:

Let \(y={y_0,y_1,...,y_i,...y_N}\) denotes the manually annotated onset positions for one of the four heart sound states while \(\hat{y}\) represents the state onsets based on the predicted states \(\hat{s}\). Assuming the tolerance parameter is \(\sigma\), the predicted segmented onset is expected to appear in the time region \(y_i-\sigma\le\hat{y_i}<{y_i+\sigma}\) and should not in the time interval \(y_i+\sigma\le\hat{y_i}<{y_{i+1}-\sigma}\). \(N_1\) and \(N_2\) would denote the counted numbers of the predicted start positions within the two time intervals. Therefore, a successful prediction happens when \(N_1=1\) and \(N_2=0\). The \(TP\), \(FP\) and \(FN\) are then counted as:
\begin{align}
  TP=TP+1, \qquad\qquad \mbox{ if } N_1>0,\\
  FP=\left\{
    \begin{aligned}
        FP+N_1-1, \qquad\qquad \mbox{ if } N_1>1,\\
        FP+N_2, \qquad\qquad \mbox{ if } N_2>0,
    \end{aligned}
    \right.\\
  FN=FN+1, \qquad\qquad \mbox{ if } N_1=0.
\end{align}
The tolerance parameter \(\sigma\) was set to 100 (ms) to evaluate different heart sound segmentation methods \cite{springer2015logistic}. The tolerance is based on the ECG R-peak detection tolerance of 150 (ms) \cite{ec571998testing}, which, as is approximately the length of the fundamental heart sounds, is shortened to 100 (ms).

Significance testing was performed using a two-sided paired t-test on the \(F_1\) scores from LEVEL-I, LEVEL-II and LEVEL-III.

\section{Results}
\begin{table*}[!ht]
    \centering
    \begin{threeparttable}
    \renewcommand{\arraystretch}{1.25}
    \caption{Total experimental results (\%) of the LR-HSMM method, the BiGRNN-based method, the TFAN-based method without the adaptive Wiener filter and the TFAN-based method with the adaptive Wiener filter (proposed) on all of the data sets from the 2016 PhysioNet/CinC Challenge. The metrics were calculated from the total number of \(TP\), \(FP\) and \(FN\) in each database. }
    \label{tab4}
    \setlength{\tabcolsep}{3mm}
    \begin{tabular}{llllllllll}
      \hline
       \multirow{2}{*} {Database} & \multirow{2}{*} {Method} & \multicolumn{5}{l}{\(F_1\) measurement for each state} & \multicolumn{3}{l}{Overall evaluation metrics} \\
      \cline{3-10}
       & & \(F_1^{S1}\) & \(F_1^{sys}\) & \(F_1^{S2}\) & \(F_1^{dia}\) & & \(Se\) & \(P_+\) & \(F_1\) \\
      \hline
      \multirow{3}{*}{Training-A\(^*\)} &LR-HSMM & \textbf{97.56} & 97.42 & 96.45 & 95.59 & & 97.37 & 97.10 & 96.75 \\
      &BiGRNN & 97.04 & 97.40 & 97.21 & 96.14 & & 97.13 & 96.50 & 97.06 \\
      &TFAN & 97.36 & 97.58 & 97.05 & 96.40 & & 97.42 & 96.78 & 97.10 \\
      &TFAN+Adaptive Wiener Filter (proposed) & 97.35 & \textbf{97.69} & \textbf{97.26} & \textbf{96.54} & & \textbf{97.49} & \textbf{96.94} & \textbf{97.21} \\
      \cline{1-10}
      \multirow{3}{*}{Training-B} &LR-HSMM & 99.43 & 99.47 & 98.64 & 98.55 & & 99.37 & 98.68 & 99.02 \\
      &BiGRNN & 93.62 & 93.90 & 91.45 & 91.69 & & 93.25 & 92.07 & 92.66 \\
      &TFAN & 98.81 & 99.16 & 98.55 & 98.56 & & 99.16 & 98.38 & 98.77 \\
      &TFAN+Adaptive Wiener Filter (proposed) & \textbf{99.65} & \textbf{99.61} & \textbf{99.29} & \textbf{99.09} & & \textbf{99.41} & \textbf{99.41} & \textbf{99.41} \\
      \cline{1-10}
      \multirow{3}{*}{Training-C} &LR-HSMM & 93.84 & 91.84 & 87.14 & 85.57 & & 88.25 & 90.97 & 89.59 \\
      &BiGRNN & 95.47 & 94.44 & 88.33 & 88.01 & & 90.91 & 92.20 & 91.55 \\
      &TFAN & 98.19 & 96.43 & \textbf{92.38} & \textbf{91.84} & & \textbf{94.09} & \textbf{95.33} & \textbf{94.71} \\
      &TFAN+Adaptive Wiener Filter (proposed) & \textbf{98.27} & \textbf{96.48} & 91.21 & 90.71 & & 93.32 & 95.03 & 94.16 \\
      \cline{1-10}
      \multirow{3}{*}{Training-D} &LR-HSMM & 96.04 & 96.04 & 93.94 & 91.69 & & 93.21 & 95.67 & 94.43 \\
      &BiGRNN & 96.88 & 96.67 & 96.97 & 95.02 & & 96.19 & 96.06 & 96.35 \\
      &TFAN & 96.37 & 96.66 & 96.80 & 95.11 & & 96.11 & 96.36 & 96.23 \\
      &TFAN+Adaptive Wiener Filter (proposed) & \textbf{96.97} & \textbf{97.66} & \textbf{97.04} & \textbf{95.53} & & \textbf{96.90} & \textbf{96.70} & \textbf{96.80}\\
      \cline{1-10}
      \multirow{3}{*}{Training-E\(^*\)} &LR-HSMM & 98.32 & 98.19 & 96.69 & 96.04 & & 96.23 & 98.41 & 97.31 \\
      &BiGRNN & 96.15 & 96.24 & 92.56 & 92.53 & & 93.26 & 95.50 & 94.37 \\
      &TFAN & 97.50 & 97.51 & 95.87 & 94.75 & & 95.48 & 97.35 & 96.41 \\
      &TFAN+Adaptive Wiener Filter (proposed) & \textbf{98.37} & \textbf{98.60} & \textbf{97.49} & \textbf{96.59} & & \textbf{97.10} & \textbf{98.43} & \textbf{97.76} \\
      \cline{1-10}
      \multirow{3}{*}{Training-F} &LR-HSMM & 90.19 & 90.51 & 87.89 & 87.45 & & 87.61 & 90.45 & 89.01 \\
      &BiGRNN & 86.49 & 86.80 & 86.31 & 86.37 & & 87.64 & 85.38 & 86.50 \\
      &TFAN & 90.69 & 90.63 & 89.64 & 88.56 & & 90.61 & 89.16 & 89.88 \\
      &TFAN+Adaptive Wiener Filter (proposed) & \textbf{92.91} & \textbf{93.29} & \textbf{92.70} & \textbf{91.30} & & \textbf{92.63} & \textbf{92.47} & \textbf{92.55} \\
      \cline{1-10}
      \multirow{3}{*}{Test-B} &LR-HSMM & \textbf{97.31} & \textbf{97.59} & \textbf{95.44} & \textbf{94.70} & & \textbf{96.60} & \textbf{95.90} & \textbf{96.25} \\
      &BiGRNN & 94.60 & 94.09 & 90.41 & 90.56 & & 93.10 & 91.72 & 92.40 \\
      &TFAN & 96.57 & 95.70 & 93.99 & 93.67 & & 95.80 & 94.16 & 94.97 \\
      &TFAN+Adaptive Wiener Filter (proposed) & 96.00 & 95.88 & 92.02 & 92.47 & & 94.51 & 93.66 & 94.09 \\
      \cline{1-10}
      \multirow{3}{*}{Test-C} &LR-HSMM & 95.60 & 95.60 & 85.06 & 83.69 & & 88.24 & 91.76 & 89.97 \\
      &BiGRNN & 97.92 & 97.68 & 95.18 & 94.00 & & 95.46 & 96.92 & 96.19 \\
      &TFAN & 98.15 & 97.56 & 95.18 & 93.76 & & 95.43 & 96.89 & 96.16 \\
      &TFAN+Adaptive Wiener Filter (proposed) & \textbf{98.45} & \textbf{98.22} & \textbf{94.89} & \textbf{94.42} & & \textbf{95.93} & \textbf{97.05} & \textbf{96.49} \\
      \cline{1-10}
      \multirow{3}{*}{Test-D} &LR-HSMM & 96.36 & 95.45 & 92.24 & 92.04 & & 93.11 & 94.90 & 94.00 \\
      &BiGRNN & 98.43 & 98.43 & 98.88 & 98.10 & & 99.11 & 97.81 & 98.45 \\
      &TFAN & 98.43 & 98.43 & 98.21 & 97.89 & & 99.00 & 97.48 & 98.24 \\
      &TFAN+Adaptive Wiener Filter (proposed) & \textbf{98.65} & \textbf{99.10} & \textbf{98.88} & \textbf{98.94} & & \textbf{99.44} & \textbf{98.35} & \textbf{98.90} \\
      \cline{1-10}
      \multirow{3}{*}{Test-E\(^*\)} &LR-HSMM & 98.02 & 98.01 & 96.59 & 95.93 & & 96.95 & 97.33 & 97.14 \\
      &BiGRNN & 98.61 & 98.59 & 96.63 & 96.79 & & 97.28 & 98.03 & 97.66 \\
      &TFAN & 99.15 & 98.98 & 98.15 & 97.87 & & 98.63 & 98.45 & 98.54 \\
      &TFAN+Adaptive Wiener Filter (proposed) & \textbf{99.22} & \textbf{99.24} & \textbf{98.11} & \textbf{98.03} & & \textbf{98.63} & \textbf{98.67} & \textbf{98.65} \\
      \cline{1-10}
      \multirow{3}{*}{Test-G} &LR-HSMM & \textbf{96.59} & \textbf{96.36} & 93.90 & 93.64 & & 94.15 & \textbf{96.10} & 95.12 \\
      &BiGRNN & 91.93 & 92.83 & 92.56 & 92.57 & & 92.80 & 92.15 & 92.47 \\
      &TFAN & 94.29 & 94.22 & 93.77 & 93.94 & & 94.71 & 93.41 & 94.06 \\
      &TFAN+Adaptive Wiener Filter (proposed) & 96.22 & 96.13 & \textbf{95.90} & \textbf{95.52} & & \textbf{96.05} & 95.84 & \textbf{95.94} \\
      \cline{1-10}
      \multirow{3}{*}{Test-I} &LR-HSMM & \textbf{99.61} & 99.08 & 92.44 & 93.60 & & 96.10 & 96.24 & 96.18 \\
      &BiGRNN & 95.63 & 96.12 & 93.92 & 94.15 & & 94.95 & 95.42 & 94.49 \\
      &TFAN & 99.52 & \textbf{99.61} & \textbf{98.08} & 97.37 & & 98.67 & \textbf{98.61} & 98.64 \\
      &TFAN+Adaptive Wiener Filter (proposed) & 99.22 & 99.39 & 97.99 & \textbf{98.19} & & \textbf{98.85} & 98.55 & \textbf{98.70} \\
      \cline{1-10}
      \multirow{3}{*}{Global Average} &LR-HSMM & 96.57 & 96.30 & 93.04 & 92.37 & & 93.93 & 95.29 & 94.56 \\
      &BiGRNN & 95.24 & 95.27 & 93.37 & 93.00 & & 94.26 & 94.15 & 94.18 \\
      &TFAN & 97.09 & 96.87 & 95.64 & 94.98 & & 96.26 & 96.03 & 96.14 \\
      &TFAN+Adaptive Wiener Filter (proposed) & \textbf{97.61} & \textbf{97.61} & \textbf{96.07} & \textbf{95.94} & & \textbf{96.69} & \textbf{96.76} & \textbf{96.72} \\
      \cline{1-10}
      \hline
    \end{tabular}
    \begin{tablenotes}
        \item[1] \(*\) in Training-A\(^*\) indicates that the 50 recordings in training set were excluded from Training-A for testing.
        \item[2] \(*\) in Training-E$^*$ and Test-E$^*$ indicates that part of original Training-E and Test-E were utilized for testing.
    \end{tablenotes}
    \end{threeparttable}
\end{table*}

 \begin{table*}[!ht]
  \centering
  \renewcommand{\arraystretch}{1.25}
  \caption{Statistical results (\%) of the LR-HSMM method, the BiGRNN-based method and the TFAN-based method among all the recordings in LEVEL-I, LEVEL-II and LEVEL-III. The performance metric means and standard errors are computed over each recording of the database respectively.}
  \label{tab5}
  \setlength{\tabcolsep}{1.5mm}
  \begin{tabular}{llllllllll}
    \hline
    \multirow{2}{*} {Database} & \multirow{2}{*} {Method} & \multicolumn{5}{l}{\(F_1\) measurement for each state} & \multicolumn{3}{l}{Overall evaluation metrics} \\
    \cline{3-10}
     & & \(F_1^{S1}\) & \(F_1^{sys}\) & \(F_1^{S2}\) & \(F_1^{dia}\) & & \(Se\) & \(P_+\) & \(F_1\) \\
    \hline
    \multirow{3}{*}{LEVEL-I} &LR-HSMM & {99.15$\pm$0.27} & {98.71$\pm$0.31} & {98.38$\pm$0.68} & {97.23$\pm$0.70} & & {97.55$\pm$0.48} & {99.26$\pm$0.36} & {98.37$\pm$0.44} \\
    & BiGRNN & 99.27$\pm$0.15 & 99.62$\pm$0.12 & 99.35$\pm$0.17 & 97.78$\pm$0.25 & & 98.53$\pm$0.15 & 99.51$\pm$0.12 & 99.01$\pm$0.13 \\
    &TFAN+Adaptive Wiener Filter (proposed) & \textbf{99.43$\pm$0.12} & \textbf{99.78$\pm$0.09} & \textbf{99.60$\pm$0.11} & \textbf{98.04$\pm$0.21} & & \textbf{98.73$\pm$0.13} & \textbf{99.71$\pm$0.09} & \textbf{99.21$\pm$0.10} \\
    \cline{1-10}
    \multirow{3}{*}{LEVEL-II} &LR-HSMM & 89.35$\pm$1.54 & 89.13$\pm$1.51 & 86.88$\pm$1.74 & 84.81$\pm$1.77 & & 86.37$\pm$1.61 & 89.49$\pm$1.42 & 87.56$\pm$1.54 \\
    & BiGRNN & 93.41$\pm$0.82 & 93.50$\pm$0.79 & 92.30$\pm$0.76 & 91.28$\pm$0.78 & & 93.17$\pm$0.68 & 92.27$\pm$0.81 & 92.63$\pm$0.73 \\
    &TFAN+Adaptive Wiener Filter (proposed) & \textbf{94.88$\pm$0.74} & \textbf{95.39$\pm$0.70} & \textbf{94.03$\pm$0.76} & \textbf{92.40$\pm$0.80} & & \textbf{94.39$\pm$0.66} & \textbf{94.06$\pm$0.77} & \textbf{94.17$\pm$0.71} \\
    \cline{1-10}
    \multirow{3}{*}{LEVEL-III} &LR-HSMM & 85.22$\pm$1.81 & 82.97$\pm$1.99 & 73.59$\pm$2.71 & 71.68$\pm$2.70 & & 76.30$\pm$2.19 & 82.13$\pm$1.82 & 78.46$\pm$2.05 \\
    & BiGRNN & 91.13$\pm$1.09 & 90.44$\pm$1.25 & 86.40$\pm$1.38 & 85.76$\pm$1.49 & & 88.62$\pm$1.22 & 88.52$\pm$1.23 & 88.45$\pm$1.20 \\
    &TFAN+Adaptive Wiener Filter (proposed) & \textbf{94.73$\pm$0.77} & \textbf{93.27$\pm$1.05} & \textbf{89.32$\pm$1.29} & \textbf{87.83$\pm$1.43} & & \textbf{90.64$\pm$1.03} & \textbf{92.12$\pm$0.99} & \textbf{91.31$\pm$1.00} \\
    \cline{1-10}
    \hline
  \end{tabular}
\end{table*}

\begin{figure*}[!ht]
  \centering
  \includegraphics[width=0.32\textwidth]{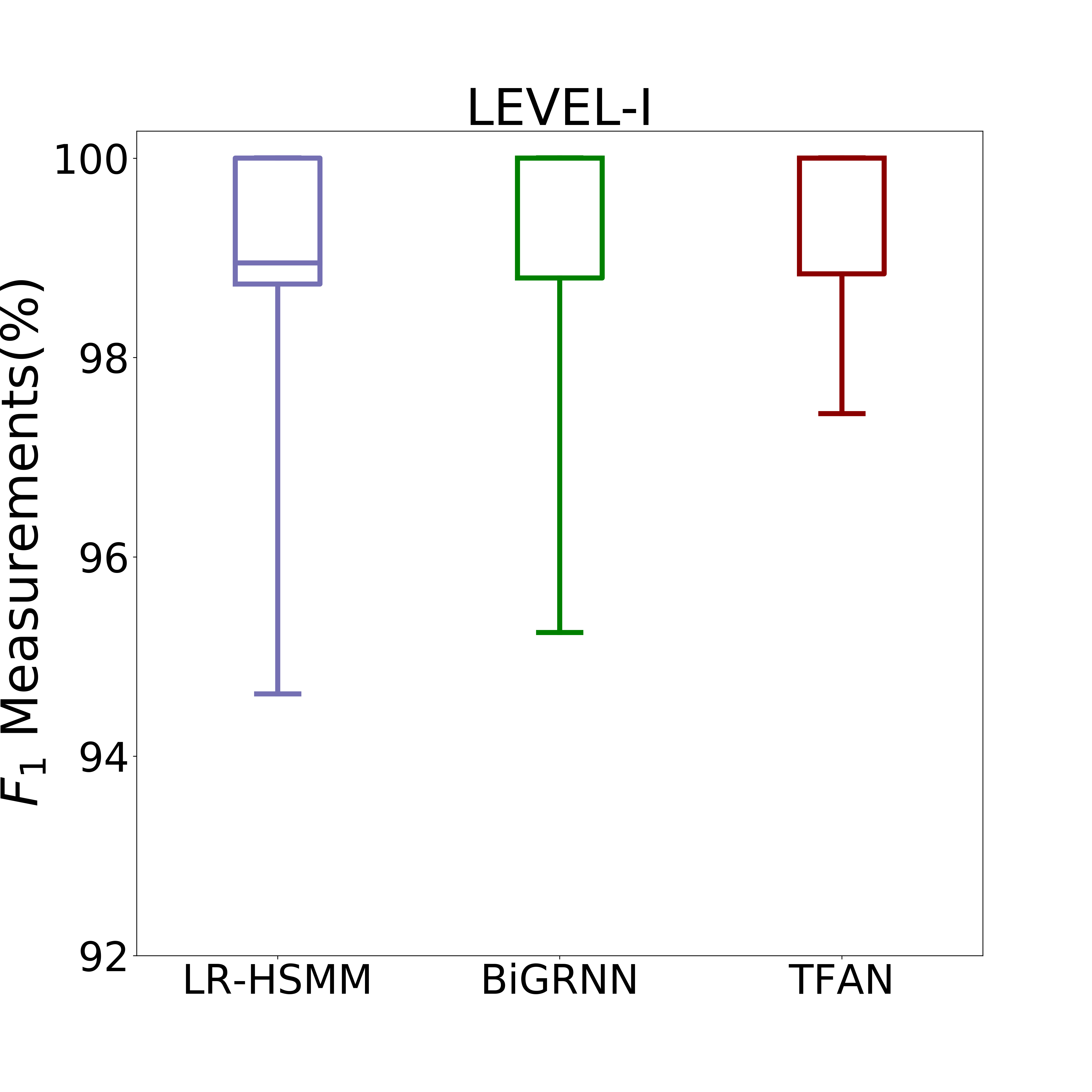}
  \includegraphics[width=0.32\textwidth]{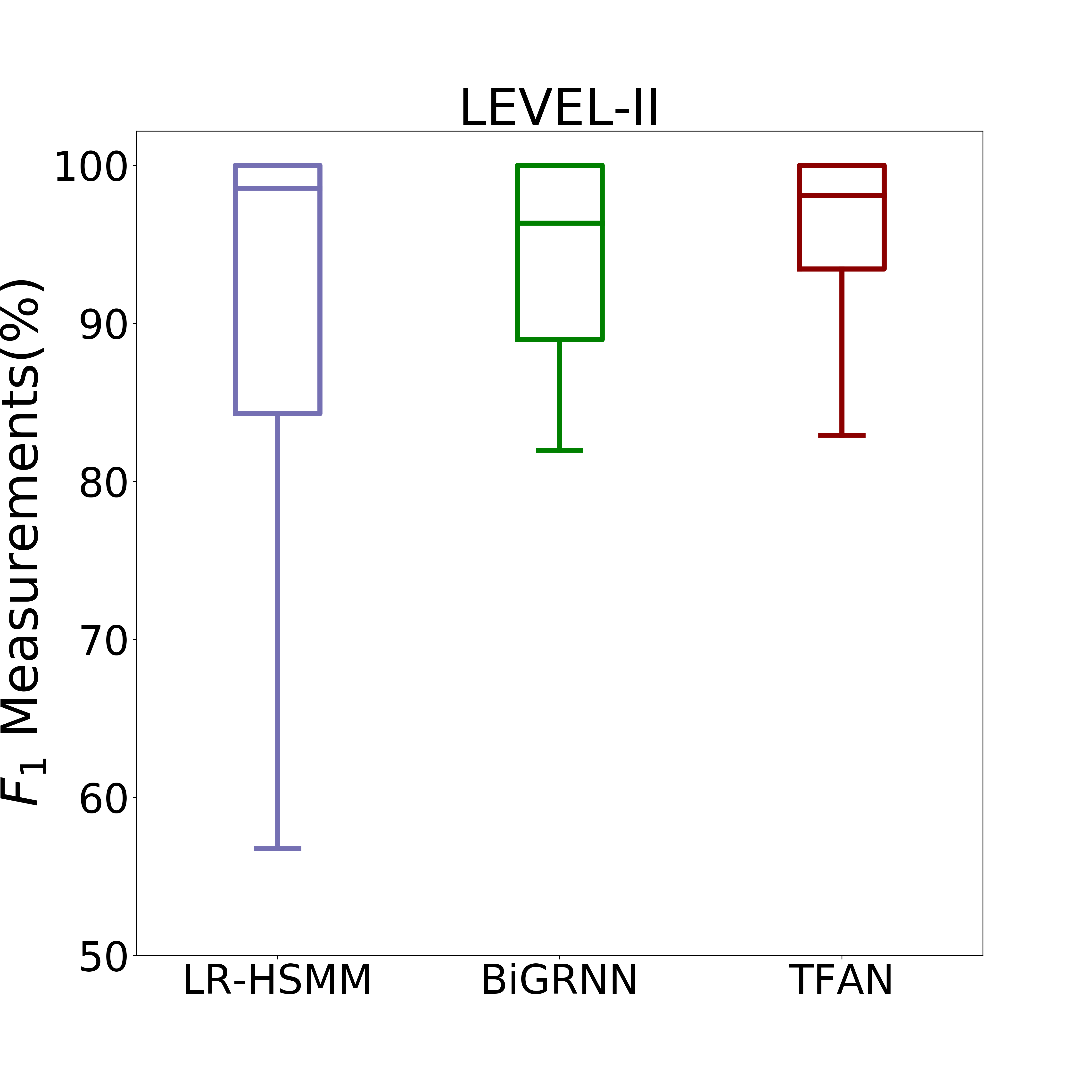}
  \includegraphics[width=0.32\textwidth]{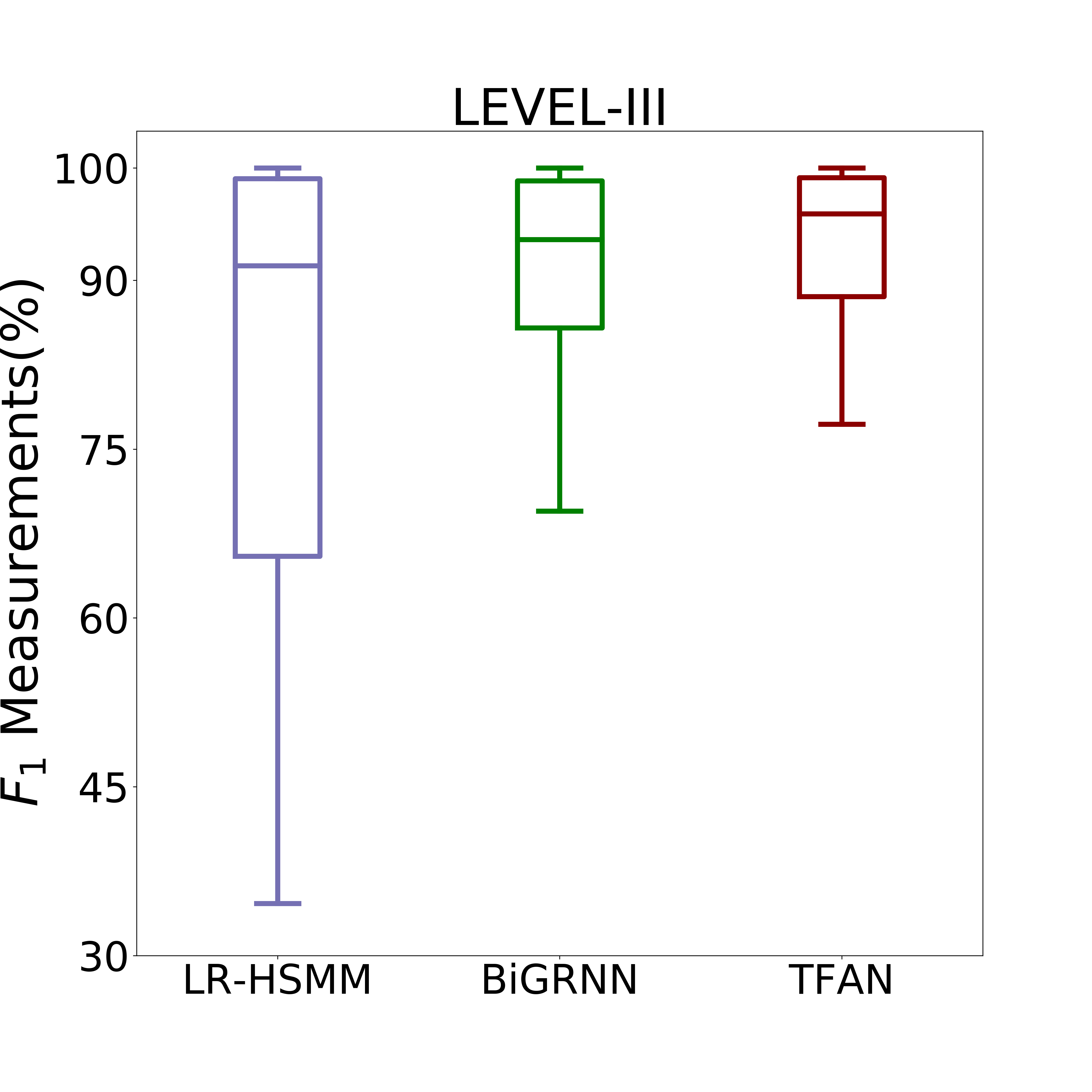}
  \caption{\(F_1\) measurements of the LR-HSMM method, the BiGRNN-based method and the TFAN-based method on databases with three levels of difficulty.}
   \label{fig8}
\end{figure*}

The gross performance results of the LR-HSMM method, the BiGRNN-based method and the TFAN-based method on all the test sets were presented in Table \ref{tab4}. The TFAN-based method was tested with and without the adaptive Wiener filter. Table \ref{tab4} illustrates the performance for the combined four states (S1, systole, S2 and diastole), as well as the \(F_1\) scores for each state separately to give an indication of performances on different states. 

The average performance of the TFAN-based method, the BiGRNN-based method and the LR-HSMM method on LEVEL-I, LEVEL-II and LEVEL-III were reported in Table \ref{tab5}.
These gross scores were calculated on a per recording basis, calculating the different metrics for each recording, then average over recordings in each of the data sets. The standard error of the averages results was also shown.

Fig.\ref{fig8} illustrated the discrepancy of the performance stability over each heart sound recording across the TFAN-based method, the BiGRNN-based method and the LR-HSMM method on LEVEL-I, LEVEL-II and LEVEL-III.

\subsubsection{Comparison with the LR-HSMM method}
According to Table \ref{tab4}, the TFAN-based method outperformed the LR-HSMM method on most of the test sets, especially on Training-C, Training-D, Training-F, Test-C, Test-D and Test-I. 
The LR-HSMM method achieved the total \(F_1\) score of \(89.59\%\) on Training-C, \(94.43\%\) on Training-D, \(89.01\%\) on Training-F, \(89.97\%\) on Test-C, \(94.00\%\) on Test-D, \(96.18\%\) on Test-I, while an enormously improvement can be seen for the TFAN-based method with the \(F_1\) score of \(94.71\%\), \(96.80\%\), \(92.55\%\), \(96.49\%\), \(98.90\%\) and \(98.70\%\) respectively. 
However, the total \(F_1\) score of the TFAN-based method on Test-B is slightly lower than LR-HSMM (\(96.25\%\)), which was \(94.97\%\) without the adaptive Wiener filter and \(94.09\%\) with the adaptive Wiener filter.

In Table \ref{tab5}, a significant improvement of performance on the TFAN-based method compared to LR-HSMM could be observed On Level-II and Level-III (\(94.17\%\) to \(87.56\%\), \(p < 0.0001\) and \(91.31\%\) to \(78.46\%\), \(p < 0.0001\)).
In comparison of standard errors, the TFAN-based method reduced the standard error by at least a factor of two comparing to the LR-HSMM method.

Fig. \ref{fig9} showed two examples of automatically segmented heart sound recordings by the TFAN-based method and the LR-HSMM method. 
Repeated mistakes happened in Fig. 9(a) and Fig. 9(b) for the LR-HSMM method when segmenting PCG signals of arrhythmia and tachycardia.

\subsubsection{Comparison with the BiRNN-based method}
According to Table \ref{tab4}, the TFAN-based method outperformed the BiGRNN-based method on the whole data sets. 
Their overall \(F_1\) scores approximated on Training-A\(^*\), Training-D, Test-C and Test-D.
Meanwhile evident improvement in performance could be observed on Training-B (\(99.4\%\) to \(92.66\%\)), Training-E\(^*\) (\(97.76\%\) to \(94.37\%\)), Training-F (\(92.55\%\) to \(86.50\%\)), Test-G (\(95.94\%\) to \(92.47\%\)) and Test-I (\(98.70\%\) to \(94.49\%\)).

Table \ref{tab5} showed that the TFAN-based method outperformed the BiGRNN-based method on LEVEL-I, LEVEL-II and LEVEL-III.
As difficulty of segmentation escalated, the average \(F_1\) scores of the TFAN-based method increased by around \(2\%\) compared to the BiGRNN-based method (\(94.17\%\) to \(92.63\%\) on LEVEL-II and \(91.31\%\) to \(88.45\%\) on LEVEL-III).
Note that the both methods showed comparable stability on each data set based on the standard errors reported in Table \ref{tab5}.

\subsubsection{Comparison of DL-based methods and the LR-HSMM method}
The BiGRNN-based method and the proposed TFAN-based method were both DL-based methods, sharing the loss function and inference function in our experiments.
In Table \ref{tab4}, the DL-based methods performed better on Training-C and Test-C compared to the LR-HSMM method.
And the both methods failed to exceed the segmentation performance of the LR-HSMM method on Test-B.
Moreover, according to Table \ref{tab5} and Fig.\ref{fig8}, the DL-based methods provided noticeable improvement to the segmentation performance on LEVEL-II and LEVEL-III.

\subsubsection{Evaluation of adaptive Wiener filter}
The introduction of the adaptive Wiener filter in the TFAN-based method  resulted in better performance on most of the test sets (except Training-C and Test-B), especially with a nearly \(3\%\) increase on Training-F.
For Training-C and Test-B, the adaptive Wiener filter caused slightly drop of around \(1\%\) on performance when segmenting S2 and diastole.

\begin{figure}[!ht]
  \centering
     \includegraphics[width=3.3in]{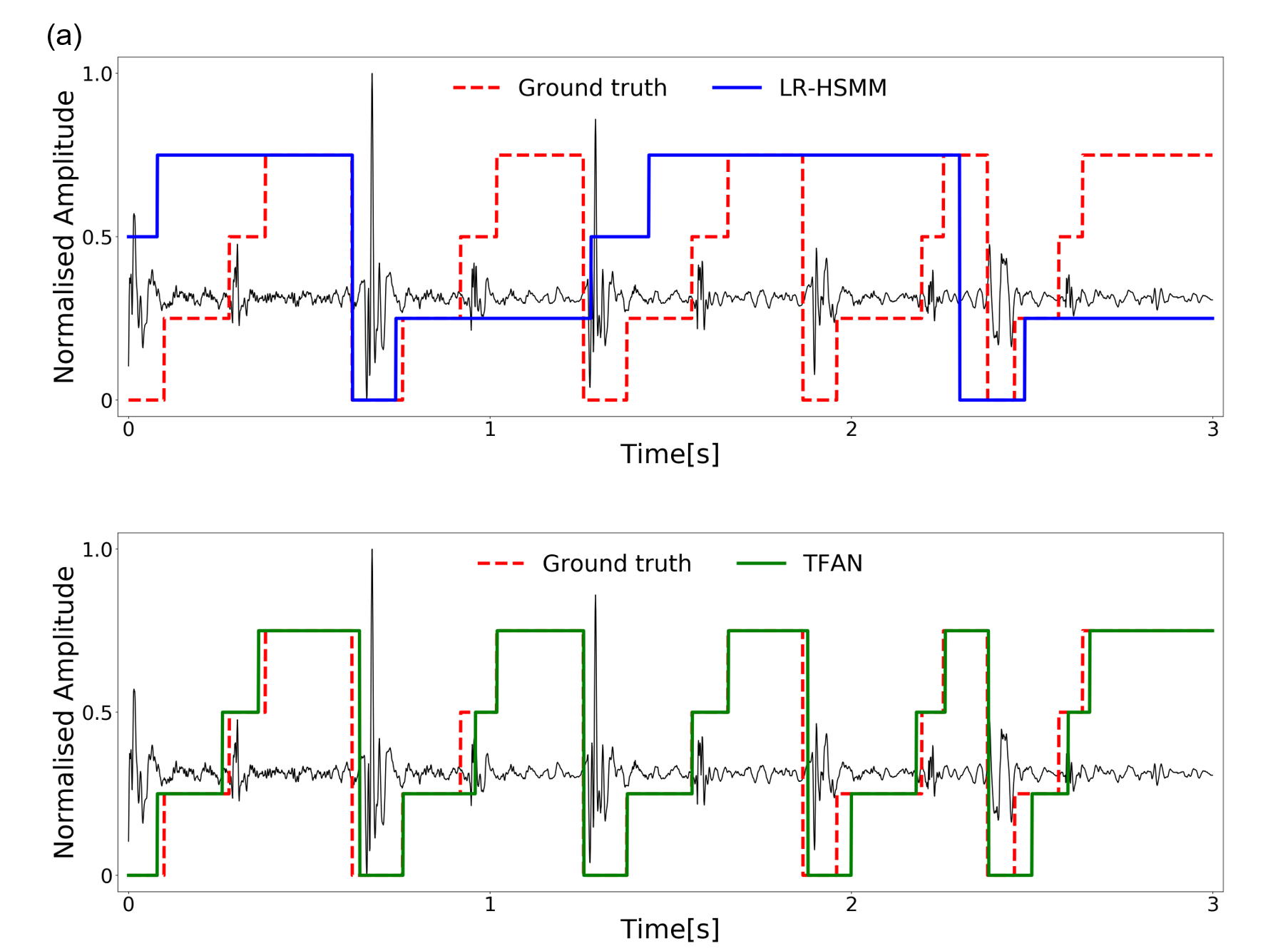}
     \includegraphics[width=3.3in]{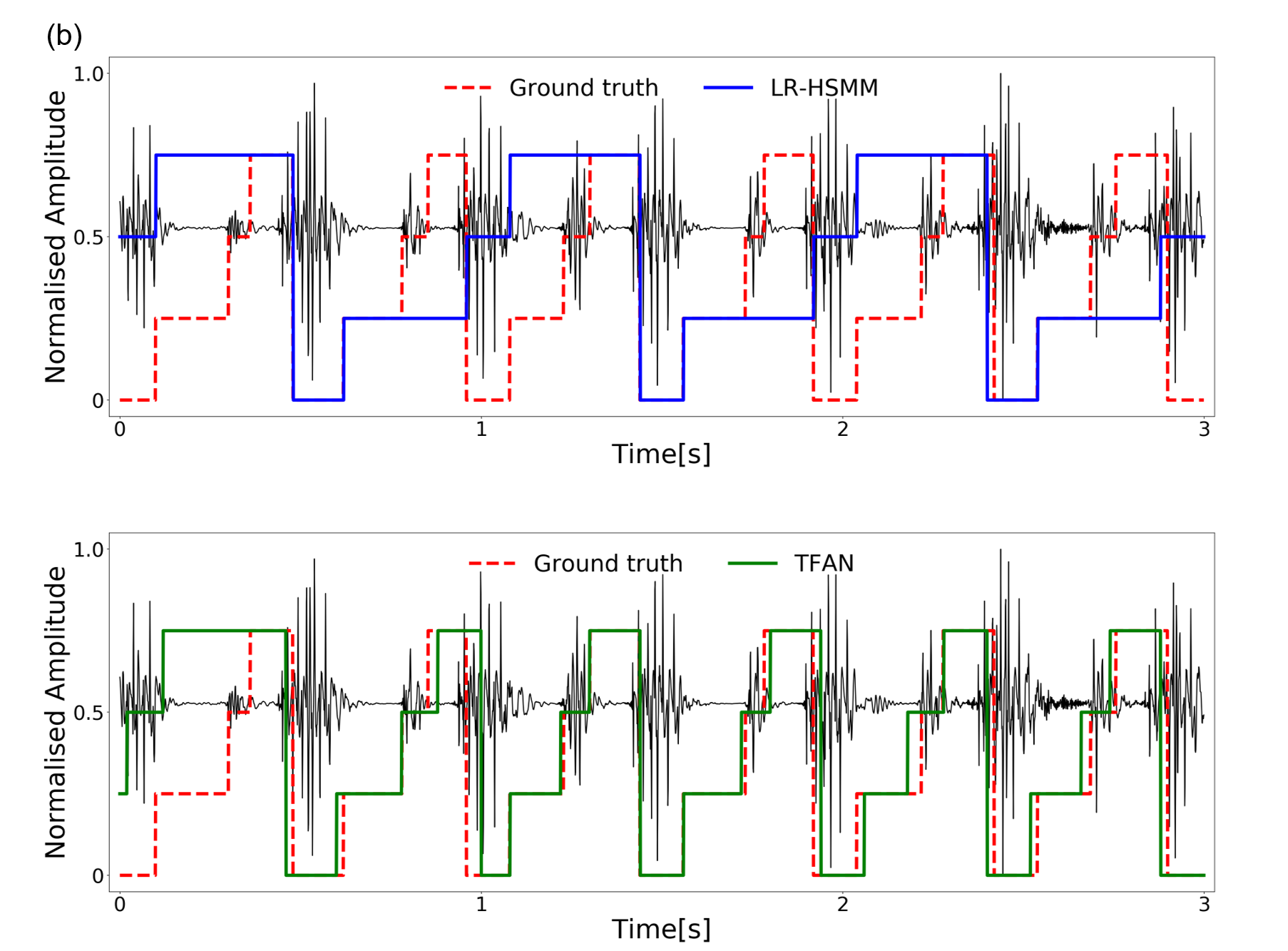}
  \caption{Example of segmented pathological PCG signal with arrhythmia (a) and tachycardia (b) using the LR-HSMM method and the TFAN-based method. The four states of the heart cycle, {S1, systole, S2, diastole}, are represented by the staircase plot with the value of {0, 0.25, 0.5, 0.75} respectively.}
  \label{fig9}
\end{figure}

\section{Discussion}
The results reported in Section V provide a comprehensive comparison between the TFAN-based method, the LR-HSMM method and the recent BiGRNN-based method. From Table. \ref{tab4} we can see that the TFAN-based method matched or outperformed the LR-HSMM method except for Test-B.
The same situation happened to another DL-based method, the BiGRNN-based method.
We hypothesize this is because Test-B is significantly different from the rest of the data, and in some way contains unusual noise or timing in the S2 and diastole periods (where the performance was most affected). 

Notably, the characteristics statistics of each data set reveal that the majority heart sound recordings in Training-B and Test-B have obscure S2 sounds (Table \ref{tab2}). 
It may be caused by contamination of noise and murmur or the stethoscopes with poor sensing performance.
Since the TFAN-based method do not use a probability distribution to constrain the duration of the states, as well the BiGRNN-based method, the methods may have failed to locate S2 unlike the LR-HSMM which can infer the most probable location even in high noise (the sensitivity-specificity trade-off).

The performances of the DL-based methods and the LR-HSMM method were both outstanding when dealing with heart sounds from patients with normal sinus rhythm. 
Therefore, the global average overall F-scores of the three methods are both in the mid 90's, with the TFAN-based method ($F_1=96.72\%$) outperforming the BiGRNN-based method ($F_1=94.18\%$) and the LR-HSMM method ($F_1=94.56\%$).
For data sets containing a certain amount of heart sound recordings with abnormal heart rhythms, such as Training-C, Training-D, Training-F and Test-C, we observe that the DL-based methods are always superior to the LR-HSMM method.
The examples in Fig. \ref{fig9} further highlight the distinctions.

From Fig. \ref{fig9}(b), we observe that although the TFAN-based method miss detects the S1 sound at the beginning, the following segmentation will be corrected in time.
Unlike the TFAN-based method, the LR-HSMM method always fails to detect events when the intervals were irregular or incompatible with the prior probabilistic distribution of the state duration. 
This is a significant result, since we are looking to diagnose abnormality and the DL-based methods (TFAN and BiGRNN) are more applicable in real-world clinical environment.

Utilization of adaptive Wiener filter in the TFAN-based method was effective in most of the situations except for Training-C and Test-B, which led to a slight drop in performance for segmenting S2 and diastole. 
The most likely reason is that the adaptive Wiener filter may attenuate the weak murmurs and disappearing S2 sounds.
This can be improved by redesigning the Wiener filter to include these specific features in the pass band. 

The improvement of performance was reinforced when the three methods were tested on the data sets with different difficulty levels (LEVEL-I, LEVEL-II and LEVEL-III). 
We refer to the results reported for LEVEL-III, with the TFAN-based method ($F_1=91.31\%$) outperforming the LR-HSMM method ($F_1=78.46\%$) and the BiGRNN-based method ($F_1=88.45\%$).
Moreover, according to Table \ref{tab5} and Fig.\ref{fig8}, we observe that the stability of the TFAN-based method and the BiGRNN-based method is superior to the LR-HSMM method in segmenting complicated heart sounds.

The BiGRNN-based method matches the TFAN-based method on a certain number of data sets with relatively higher signal quality.
But obviously, the generalization of the BiGRNN-based method is inferior to the TFAN-based method.
In comparison of model size, the weight file of TFAN is 2.3Mb with 290,453 parameters, superior to BiGRNN model, which is 8.1Mb with 1,016,805 parameters. 
Considering the both methods shared the proposed loss function and the designed dynamic inference approach, the advantage of TFAN over BiGRNN is in model structure.

The proposed framing module in TFAN slices the feature map into frames after the encoder (see Section III-B).
Then the decoder implicitly learns the conditional probability distribution of state given encoded feature matrix for each time step.
Comparing to BiGRNN, this structure enables TFAN to be more flexible in learning the decision boundaries between distributions of different heart sound states and brings the advantage of dealing with out-of-distribution data.
Therefore, although the training set was limited into 50 recordings, the TFAN-based method still achieved the best performance.

The inner framing operation is also equivalent to incorporating the feature transformation of the signal during the model learning process.
On the premise of ensuring the time resolution as much as possible, the feature expression dimension in each frame is improved.
Unlike non-adaptive static feature extraction methods, such as envelop filters and spectrogram transform, this structure makes the model capable of capturing the features of the inter-state variability and the state transition information dynamically.
This results in a high sensitivity for detecting the onsets and offsets of S1 and S2 precisely and reduces errors introduced by other heart sounds (e.g. S3) and noise.
Moreover, the proposed TFAN-based method does not introduce the current error information into subsequent calculations for identifying the S1 and S2 in the next cycle. 

However, there are two key limitations to the TFAN-based method. Firstly, the TFAN-based method was prone to missing weak or disappearing S2 sounds and identified the subsequent S1 sound as S2 in such cases. Secondly, the TFAN-based method tended to falsely identify some brief or transient noise as S1 or S2 sounds if the noises were similar to S1 or S2. This is basically an inherent problem of any classification technique, although with enough data we expect to be able to remove such events that appear at implausible times in the sequence of states, considering all pathological states. 

\section{Conclusion}
This paper proposed a novel method for heart sound segmentation of S1, systole, S2 and diastole. The method built up a frame-level feature classifier for the four components by an original temporal framing network. The study was focused on how to incorporate the state transition information into algorithm without using HMMs. The introduction of state transition loss and dynamic inference effectively addressed the problem within one model. Moreover, the TFAN-based method did not require explicit modeling of timing and was therefore able to generalize to arrhythmia and other high variability recordings more effectively than the current state of the art. Even though the training set was restricted to a small database with 50 single-source recordings randomly selected from Training-A, it was noted that the TFAN-based method provided a substantial improvement, particularly for more difficult cases, and on data sets not represented in the public training data.
Future work will examine how increasing the number of training patterns and modeling the distribution of latent space to improve the performance.
However, we note that the more data we use, the more we must use lower quality data, or make enormous effort to improve the labels. 

Further work is also required to understand how this approach will provide improved diagnostic performance, although it is logical to assume better segmentation will lead to improved diagnostics.

\appendices

\ifCLASSOPTIONcaptionsoff
  \newpage
\fi

\section*{Acknowledgment}
This work was supported by the Distinguished Young Scholars of Jiangsu Province under grant BK20190014, the National Natural Science Foundation of China under grant 81871444 and the Primary Research \& Development Plan of Jiangsu Province under grant BE2017735. This work was also supported by the National Institutes of Health-sponsored Research Resource for Complex Physiologic Signals (www.physionet.org) (R01GM104987). The content of this article is solely the responsibility of the authors and does not necessarily represent the official views of the National Institutes of Health.

\bibliographystyle{IEEEtran}
\bibliography{references}

\end{document}